\documentclass[prd,aps,12 pt, nofootinbib, preprintnumbers]{revtex4-2}
\pdfoutput=1
\usepackage{graphicx}
\usepackage{amssymb}
\usepackage{mathtools}
\usepackage{epstopdf}
\usepackage{color}
\usepackage{bbold}
\usepackage{float}
\usepackage[title]{appendix}
\usepackage[colorlinks, citecolor=blue, linkcolor=black, urlcolor=purple]{hyperref}
\usepackage{amsfonts}
\usepackage{amsmath}
\usepackage{setspace}
\usepackage{epsfig}
\usepackage[table,xcdraw]{xcolor}
\usepackage{tabularx}
\usepackage{empheq}
\usepackage{mathrsfs}
\usepackage{xcolor}
\usepackage{ulem}
\usepackage{slashed}
\usepackage{multirow}
\usepackage{braket}
\usepackage{dsfont}
\usepackage{upgreek}
\usepackage{bigints}
\usepackage{ stmaryrd }
\usepackage{comment}
\usepackage{caption}
\usepackage{subcaption}
\usepackage{geometry}

\usepackage{units}

\newcommand{\be}{\begin{eqnarray}}
\newcommand{\ee}{\end{eqnarray}}
\newcommand{\bea}{\begin{eqnarray}}
\newcommand{\eea}{\end{eqnarray}}

\newcommand{\beq}{\begin{equation}}
\newcommand{\eeq}{\end{equation}}

\definecolor{bluDT}{cmyk}{1,0.5,0,0.3}

\definecolor{darkblue}{rgb}{0.2,0.2,0.9}
\definecolor{colorRTD}{rgb}{.2,.2,.7}

\definecolor{colorHD}{rgb}{.2,0.9,.0.9}

\usepackage{soul}

\begin{document}

%\preprint{CERN-TH-2021-138}

%%%%%%%%%%%%%%%%%%%%%%%%%%%%%%%%%%%%%%%%%%%%%%%%%%%%%%%
\title{\LARGE A functional treatment of small instanton-induced axion potentials
}
%%%%%%%%%%%%%%%%%%%%%%%%%%%%%%%%%%%%%%%%%%%%%%%%%%%%%%%

\author{Pablo Sesma}
\affiliation{Université Paris-Saclay, CEA, CNRS, Institut de Physique Théorique, 91191, Gif-sur-Yvette, France}

%%%%%%%%%%%%%%%%%%%%%%%%%%%%%%%%%%%%%%%%%%%%%%%%%%%%%%%
%\date{}
\begin{abstract}
\mbox{} \\ 
\begin{center}
{\bf Abstract}
\end{center}
We present a functional method to perform complete one-instanton calculations of the axion potential. This is done for an $SU(N)$ gauge theory with a matter content in any representation of the gauge group. This type of computation requires the expression of the fermion zero modes of the theory. We construct them for all representations of $SU(2)$, which serve as building blocks for obtaining the fermion zero modes for arbitrary representations of $SU(N)$. The method is applied to the Minimal Supersymmetric $SU(5)$ model and its low-energy counterpart, the Minimal Supersymmetric Standard Model extended with two color triplets.
\end{abstract}

%%%%%%%%%%%%%%%%%%%%%%%%%%%%%%%%%%%%%%%%%%%%%%%%%%%%%%%
\maketitle
\onecolumngrid
%%%%%%%%%%%%%%%%%%%%%%%%%%%%%%%%%%%%%%%%%%%%%%%%%%%%%%%
\newpage
\tableofcontents
\newpage
%%%%%%%%%%%%%%%%%%%%%%%%%%%%%%%%%%%%%%%%%%%%%%%%%%%%%%%
\section{Introduction}

The Strong CP problem arises from the absence of observed CP violation in Quantum Chromodynamics (QCD), despite QCD allowing CP-violating terms in its Lagrangian. In the Standard Model, the QCD Lagrangian includes the operator $\text{Tr}[G_{\mu\nu}\widetilde{G}^{\mu\nu}]$ with a coefficient proportional to $\bar{\theta}=\theta-\text{arg }\text{det}\left[Y_u Y_d\right]$, where $G_{\mu\nu}$ is the gluon field strength, $\widetilde{G}_{\mu\nu}$ is its Hodge dual, $\theta$ is the QCD vacuum angle and $Y_{u,d}$ are the up- and down-type Yukawa matrices. This term violates the combined charge conjugation (C) and parity (P) symmetries of the theory. As quarks and gluons confine into hadrons at low energies, this operator can influence hadron physics \cite{Hook:2018dlk}.
Experimental bounds, especially those related to the neutron electric dipole moment, impose a stringent constraint on $\bar{\theta}$, requiring it to be smaller than $10^{-10}$ \cite{Abel:2020pzs}. This extreme fine-tuning of $\bar{\theta}$ remains unexplained, creating the well-known Strong CP problem.

The QCD axion, introduced via the Peccei-Quinn (PQ) mechanism \cite{Peccei:1977hh,Peccei:1977ur,Weinberg:1977ma,Wilczek:1977pj}, offers the most compelling solution to the Strong CP problem and represents one of the most promising scenarios for physics beyond the Standard Model. In this framework, the axion, with a potential generated by QCD strong dynamics, relaxes to a value that exactly cancels $\bar{\theta}$, restoring CP symmetry in the strong interactions. Moreover, axions not only solve the Strong CP problem but also emerge as promising candidates for dark matter, motivating their ongoing investigation in both theoretical and experimental contexts \cite{Preskill:1982cy,Abbott:1982af,Dine:1982ah}. 

However, the PQ mechanism is confronted with a significant issue. The QCD axion arises from a symmetry that must remain nearly exact to solve the Strong CP problem. Yet, even very high-dimensional operators, albeit suppressed by the Planck scale, can misalign the axion potential, preventing it from dynamically relaxing the axion to a value that cancels the parameter $\bar{\theta}$. One way to bypass this issue is by increasing the scale of the axion potential, since a heavier axion is less sensitive to the quality problem, provided there are no new sources of CP violation at high energies that could influence the axion potential through small instanton effects \cite{Csaki:2023ziz,Bedi:2024wqg}.
Both this challenge and the phenomenological motivations surrounding axion detection have driven extensive theoretical efforts to deepen our understanding of the axion potential, which is closely tied to the dynamics of non-perturbative QCD \cite{Witten:1978bc,Veneziano:1979ec,Witten:1979vv}. As a result, numerous scenarios have been proposed to increase the axion mass \cite{Holdom:1982ex,Holdom:1985vx,Flynn:1987rs}, many of which rely on instanton-based mechanisms.

Instantons are non-perturbative solutions to the Euclidean Yang-Mills equations in non-Abelian gauge theories \cite{Belavin:1975fg,tHooft:1976rip,tHooft:1976snw}, representing tunneling events between distinct vacuum states characterized by different topological charges. When instantons were initially discovered, there were hopes that they could provide analytical insights into confinement in strongly coupled theories, as they introduce calculable non-perturbative effects in the path integral. However, being rooted in semiclassical methods, these approaches have limitations, as they rely on a small gauge coupling $g$, preventing a full understanding of confinement \cite{Callan:1977gz}.

Nevertheless, instanton calculus quickly found applications in model building. At high energies, $E\gg \Lambda_{\rm QCD}$, small instantons, with size $\rho\sim E^{-1}$, generate calculable semiclassical contributions to the effective axion potential, proportional to $e^{-8\pi^2/g^2(1/\rho)}$. While small-instanton contributions are generally anticipated to be significantly smaller than those from QCD strong dynamics, enhancements can occur under certain conditions. For instance, by modifying the running of the QCD coupling \cite{Holdom:1982ex,Holdom:1985vx,Flynn:1987rs}, it is possible to re-establish strong coupling in the UV while still maintaining the semiclassical approximation. Additionally, embedding QCD in a higher-dimensional theory—such as a 5D framework \cite{Poppitz:2002ac,Gherghetta:2020keg}—can enhance small instanton contributions, potentially making them the dominant source of the axion mass. Since these methods rely on the contributions of instantons to the axion potential, it is essential to develop robust tools that extend instanton calculus to theories with gauge groups more complex than $SU(3)$ and with more exotic matter content.

In this paper, we present a method for performing one-instanton computations of the effective axion potential in $SU(N)$ gauge theories with matter fields in any representation of the gauge group. We construct the generating functional of the theory and apply standard functional perturbative methods to evaluate diagrams in the interacting theory. This functional method not only captures all $\mathcal{O}(1)$ factors but also enables tractable calculations in more intricate scenarios, such as when scalar fields charged under the instanton's gauge group propagate to close fermion legs associated with fermion zero modes. Additionally, we construct the explicit form of the fermion zero modes for arbitrary representations of $SU(2)$, which serve as building blocks for zero modes in any representation of $SU(N)$ and other gauge groups containing $SU(2)$.
This paper aims to establish a solid foundation for calculating instanton contributions to the axion potential, offering a transparent framework for practical applications.

 The paper is organized as follows: in Section \ref{Instantons, the theta-vacuum and the axion potential}, we briefly review the BPST instanton solution and its embedding into $SU(N)$. Given that instantons mediate tunneling effects, we express the energy density of the $\theta$-vacuum, responsible for the lowering of the potential barrier, through a path integral. This expression involves a multi-instanton computation, which we simplify using the dilute instanton gas approximation, reducing it to a single-instanton calculation. From this framework, we derive the effective axion potential. In Section \ref{The one-instanton generating functional}, we introduce the method for calculating these contributions from single instantons. Instead of directly computing the vacuum-to-vacuum amplitude in the instanton background, we construct the generating functional of the theory in such a background, enabling a more tractable approach to handle interactions. To determine the instanton contributions to the axion potential, we compute vacuum diagrams, which require closing the external fermion lines associated with fermion zero modes. In Section \ref{Fermion zero modes} , we derive these zero modes for any representation of $SU(2)$ and provide a general procedure to extend this result to arbitrary representations of $SU(N)$. This method is applied in Section \ref{Examples} to the Minimal Supersymmetric Standard Model (MSSM) extended with color triplets $T_{u,d}$ which emerge after the spontaneous symmetry breaking of $SU(5)$. Ultimately, we carry out the computation in the Minimal Supersymmetric $SU(5)$ Grand Unified Theory (GUT), demonstrating that in this case small instantons are significantly suppressed.
We conclude in Section \ref{sec:conclusion}.
\newpage

%%%%%%%%%%%%%%%%%%%%%%%%%%%%%%%%%%%%%%%%%%%%%%%%%%%%%%%
\section{Instantons, the $\theta$-vacuum and the axion potential}\label{Instantons, the theta-vacuum and the axion potential}

\subsection{Lightning review of the BPST instanton solution}

Instantons are finite action solutions to the classical Euclidean Yang-Mills equations. As a result, they satisfy the self-duality equations $F_{\mu\nu}=\widetilde{F}_{\mu\nu}=\frac{1}{2}\varepsilon_{\mu\nu\rho\sigma}F_{\rho\sigma}$ and asymptotically approach pure gauge $A_{\mu}(|x|^2\rightarrow \infty)=i U^{-1}\partial_{\mu} U$, for some $U\in SU(2)$. Such solutions can be classified by an integer number called the Pontryagin index $n$ given by
\begin{equation}
    n=\frac{1}{16\pi^2}\int d^4x \text{Tr}\left[F_{\mu\nu}\widetilde{F}_{\mu\nu}\right]\, .
    \label{instanton winding number}
\end{equation}
The BPST instanton solution \cite{Belavin:1975fg} corresponds to $n=1$ and is given in the \textit{regular gauge} by
\begin{equation}
    A_{\mu}^{SU(2)}(x)=2\eta_{a\mu\nu}U(\vec{\theta})T^{a}U^{-1}(\vec{\theta})\frac{(x-x_0)_{\nu}}{(x-x_0)^2+\rho^2}\, ,\qquad U\in SU(2)\, .
\end{equation}
where $\eta_{a\mu\nu}$ is known as the 't Hooft symbol and is defined in Appendix \ref{Angular momenta of the problem}.
This solution is characterized by the set of instanton collective coordinates: the location of its center $x_0^{\mu}$, its size $\rho$ and its orientation within the $SU(2)$ gauge group parametrized by the three angles $\vec{\theta}$. Therefore, the BPST instanton solution is parametrized by $4+1+3=8$ collective coordinates.
\newline

In this paper, we focus on instanton-induced axion potentials within $SU(N)$ theories. As such, we embed the instanton solution into $SU(N)$ using the minimal embedding framework, where the instanton is placed in the upper-left corner of the fundamental representation of $SU(N)$. However, this is not the most general $SU(N)$ instanton solution, as different embeddings are possible. To account for all possible embeddings, we apply to this solution elements of $SU(N)$ that generate new configurations. These elements belong to the coset\footnote{We follow the notation $S\left[U(p)\times U(q)\right]$ from \cite{Tong:2005un}, indicating the omission of the overall central $U(1) \subset U(p)\times U(q)$.} $SU(N)/S\left[U(N-2)\times U(2)\right]$, with dimension $4N-5$, where $T_N=S\left[U(N-2)\times U(2)\right]$ is referred to as the stability group \cite{Bernard:1979qt}. Thus, the embedded instanton solution takes the form 
\begin{equation} 
A_{\mu}^{SU(N)}(x)=2\eta_{a\mu\nu}U(\Omega)T^{a}U^{-1}(\Omega)\frac{(x-x_0)_{\nu}}{(x-x_0)^2+\rho^2}\, , \qquad U\in SU(N)/{T_N}
\, ,
\label{instanton solution} 
\end{equation}
where for $a=1,2,3$, the $T^{a}$'s are the Pauli matrices embedded in the upper-left corner of the fundamental representation of $SU(N)$, while all other generators are zero.
In this case the number of collective coordinates is increased to $4+1+4N-5=4N$.

\subsection{Instantons as tunneling solutions and the energy of the $\theta$-vacuum}

The existence of instanton solutions implies the presence of distinct, inequivalent classical ground states, between which instantons mediate quantum tunneling. These states are labeled by the Pontryagin index $n$, given in Eq. $\eqref{instanton winding number}$. Since all degenerate vacua $\ket{n}$ can be reached through transition amplitudes, the true vacuum must be a superposition of all $\ket{n}$ vacua \cite{Callan:1977gz,Shifman:2012zz}. The true $\theta$-vacuum is defined to be gauge invariant under usual gauge transformations, and invariant up to an overall phase under topologically non-trivial gauge transformations\footnote{These are also known as large gauge transformations, denoted by $\Omega_m$, corresponding to a winding number $m$. Such transformations map the vacuum state $\ket{n}$ to $\Omega_m\ket{n}=\ket{n+m}$.}. It is constructed from the $\ket{n}$-vacua as
\begin{equation} 
\ket{\theta}=\sum_{n=-\infty}^{+\infty}e^{in\theta}\ket{n}\, , 
\end{equation}
We are interested in the energy density of the $\theta$-vacuum. This can be obtained from the transition amplitude
\begin{align}
\left\langle \theta \left| e^{- H T}\right| \theta\right\rangle=&\sum_{n,n'}e^{-i(n'-n)\theta}\left\langle n' \left| e^{- H T}\right| n\right\rangle=\sum_{n,n'}e^{-i(n'-n)\theta}\left\langle n'-n \left| e^{-HT}\right|0\right\rangle\, ,
\end{align}
where we used that the Hamiltonian commutes with topologically non-trivial gauge transformations. In the infinite volume limit, this transition amplitude has a Euclidean functional integral representation of the form
\begin{equation}
\lim_{V_4\rightarrow\infty}\left\langle \theta \left| e^{-H T}\right| \theta\right\rangle=\sum_{n,n'}\int \mathcal{D}A_{n'-n}\exp\left[-\int d^4x \left(\frac{1}{2g^2}\text{Tr}\left[F_{\mu\nu}F_{\mu\nu} \right]+i\frac{\theta}{16\pi^2}\text{Tr}\left[F_{\mu\nu}\widetilde{F}_{\mu\nu} \right] \right)\right]\, ,
\end{equation}
where the subscript $n'-n$ in the measure indicates that any configurations with winding number $n'-n$ are to be included, not just solutions to the equations of motion, as required by locality and unitarity \cite{Weinberg:1996kr,Reece:2023czb,Shifman:2012zz}.
\newline 

As the four dimensional volume goes to infinity, the energy density of the $\theta$-vacuum can be extracted as
\begin{equation}
e^{-E(\theta)V_4}=\lim_{V_4\rightarrow\infty}\sum_{n',n}e^{-i(n'-n)\theta}\left\langle n'-n \left| e^{-HT}\right|0\right\rangle\, .
\end{equation}
In the dilute instanton gas approximation \citep{Callan:1977gz,Coleman:1985rnk}, the $\theta$-vacuum is built from successive transitions between different $\ket{n}$ states, driven by an arbitrary number of well-separated, non-interacting instantons and anti-instantons. To construct the $\theta$-vacuum energy density, we sum over all these contributions while carefully avoiding multiple counting
\begin{equation}
e^{-E(\theta)V_4}\simeq\sum_{n_+,n_-=0}^{+\infty} \frac{\left[e^{-i\theta}\left\langle 1 \left| e^{-HT}\right|0\right\rangle\right]^{n_+}}{n_+!}\frac{\left[e^{i\theta}\left\langle -1 \left| e^{-HT}\right|0\right\rangle\right]^{n_-}}{n_-!}=e^{Z_{SU(N)}+\text{h.c.}}\, ,
\label{energy of theta vacuum}
\end{equation}
where $n_+$ and $n_-$ are the number of instantons and anti-instantons and we introduced the notation 
\begin{equation}
Z_{SU(N)}=e^{-i\theta}\left\langle 1 \left| e^{-HT}\right|0\right\rangle\, ,
\end{equation}
for the vacuum-to-vacuum amplitude in the background of a single instanton. Thus, in this approximation we have reduced a multi-instanton computation to a single instanton one.  The matrix element $Z_{SU(N)}$ can be represented as a functional integral. In the weak coupling limit, where $g^2\hbar \ll 1$, the Euclidean functional integral is evaluated using the saddle-point approximation. This approach requires a solution to the Euclidean equations of motion with appropriate boundary conditions. Instantons are such solutions, as they describe the vacuum tunneling between $\ket{n}$ and $\ket{n+1}$.

\subsection{The axion potential}

The QCD axion is the pseudo-Nambu Goldstone boson of the anomalous Peccei-Quinn (PQ) symmetry, $U(1)_{\rm PQ}$ \cite{Peccei:1977hh,Peccei:1977ur,Weinberg:1977ma,Wilczek:1977pj}. The $U(1)_{\rm PQ}$-$\rm QCD$-$\rm QCD$ chiral anomaly induces a coupling of the axion to the QCD field strength of the form
\begin{equation}
    \mathcal{L}_a=i\left(\theta+\frac{a}{f_a}\right)\frac{1}{16\pi^2}\text{Tr}\left[G_{\mu\nu}\widetilde{G}_{\mu\nu}\right]\, .
\end{equation}
The PQ symmetry being non-linearly realized, it acts on the axion as a continuous shift symmetry $a/f_a\longmapsto a/f_a+\alpha$. This is a symmetry because $\text{Tr}\left[G_{\mu\nu}\widetilde{G}_{\mu\nu}\right]$ is a total derivative, however in the presence of an instanton, the situation changes significantly. In this case, given Eq. $\eqref{instanton winding number}$, the continuous shift symmetry is broken down to a discrete shift symmetry, such that $a/f_a\longmapsto a/f_a+2\pi k$, for $k\in \mathds{Z}$.

Given that the presence of instantons explicitly breaks the continuous shift symmetry of the axion, they induce a potential for it.
This potential is constructed from Eq. $\eqref{energy of theta vacuum}$ by treating the axion as a constant background field \textit{à la} Coleman-Weinberg \cite{Coleman:1973jx}. This results in an effective axion potential of the form \cite{Csaki:2023yas}
\begin{equation}
  -\int d^4x~ V(a)\simeq\lim_{V_4\rightarrow\infty}e^{-i\frac{a}{f_a}}Z_{SU(N)}+\text{h.c.}\, .
\end{equation}
The purpose of this paper is to present a method to compute the instanton-induced effective axion potential in an $SU(N)$ gauge theory with fermions and scalars in any representation of the gauge group. Interactions between these particles play a crucial role to obtain a non-zero result. In the next section we provide a treatment of those interactions using functional methods.

%%%%%%%%%%%%%%%%%%%%%%%%%%%%%%%%%%%%%%%%%%%%%%%%%%%%%%%

%%%%%%%%%%%%%%%%%%%%%%%%%%%%%%%%%%%%%%%%%%%%%%%%%%%%%%%
\section{The one-instanton generating functional} \label{The one-instanton generating functional}

\subsection{Set-up}

In the following we consider an $SU(N)$ gauge theory with $\mathcal{S}$ complex scalar fields $\phi_i$ in the representation $\textbf{R}_{s}$, $s=1,\cdots , \mathcal{S}$ and $\mathcal{F}$ massless Weyl fermions in the representation $\textbf{R}_{f}$ of $SU(N)$, $f=1,\cdots , \mathcal{F}$. To write the Euclidean action for this theory we follow the conventions of \cite{Shifman:2012zz,Csaki:2019vte} and we decompose the action as
\begin{equation}
S_E=S_A+S_{\phi}+S_{\psi}\, ,
\end{equation}
where
\begin{align}
S_A=&\int d^4x \left(\frac{1}{2g^2}\text{Tr}\left[F_{\mu\nu}F_{\mu\nu} \right]+i\frac{\theta}{16\pi^2}\text{Tr}\left[F_{\mu\nu}\widetilde{F}_{\mu\nu} \right]+\mathcal{L}_{\rm ghost}(c,\bar{c})\right)\, ,\\
S_{\phi}=& \int d^4x \left[ (D_{\mu}\phi)^{\dagger}(D_{\mu}\phi)+V(\phi) \right]\, ,\\
S_{\psi}=& \int d^4x ~ i\psi^{\dagger}\bar{\sigma}_{\mu}D_{\mu}\psi=\int d^4x~ i\psi\sigma_{\mu}D_{\mu}\psi^{\dagger}\, ,
\end{align}
where $D_{\mu}=\partial_{\mu}-iA_{\mu}^{a}T^{a}(\textbf{R})$ is the covariant derivative acting on a field in the representation $\textbf{R}$ of $SU(N)$.
We aim to compute the vacuum-to-vacuum amplitude, where two vacua, $\ket{n}$ and $\ket{n+1}$, are interpolated by a single instanton. As mentioned in the previous section, this tunneling transition is evaluated using a semiclassical approximation, where the expansion is performed around the instanton solution.
To perform a background field expansion around the classical instanton solution $A_{\mu}^{SU(N)}$, with $\phi_i=0$ and $\psi_i=0$ for all scalars and fermions, we decompose the gauge field into a classical background field given by the instanton solution $A_{\mu}^{SU(N)}$ and a fluctuating quantum field $\mathcal{A}_{\mu}$. The remaining fields in the action are treated as quantum fluctuations around a zero background. Therefore, we compute \cite{Csaki:2019vte}
\begin{equation}
Z_{SU(N)}=e^{-i\theta}\mathcal{N}\int\mathcal{D}\mathcal{A}_{\mu}\mathcal{D}c\mathcal{D}\bar{c}\left[\mathcal{D}\psi \right]\left[\mathcal{D}\psi^{\dagger} \right]\left[\mathcal{D}\phi \right]\left[\mathcal{D}\phi^{\dagger} \right]e^{-S_E}\, .
\end{equation}
The normalization factor $\mathcal{N}$ is chosen such that the vacuum-to-vacuum amplitude in the absence of an instanton background is normalized to $1$, ensuring the vacuum state has a norm of $1$. To achieve this, we divide $Z_{SU(N)}$ by the corresponding expression expanded around the trivial background. $\mathcal{N}$ also includes the Pauli-Villars sector, which is used to regularize and renormalize the theory. Additionally, $\left[\mathcal{D}\upphi\right]$ denotes the integration measure over all $\upphi$-type fields in the theory, and the expanded action is
\begin{equation}
S_E=\frac{8\pi^2}{g^2}+\int d^4x \left[\frac{1}{2}\mathcal{A}_{\mu}^{a}\left(\mathcal{M}_{A}\right)^{ab}_{\mu\nu}\mathcal{A}_{\nu}^{b}+\bar{c}^{a}\left(\mathcal{M}_{\rm ghost}\right)^{ab}c^{b}+\phi^{\dagger}\mathcal{M}_{\phi}\phi+\psi^{\dagger}\mathcal{M}_{\psi}\psi \right]\, .
\label{expanded action}
\end{equation}
The first term corresponds to the action of the background instanton and the operators in the bracket are listed in Appendix \ref{Instanton density}. The next step is simply a functional Gaussian integration over the quantum fluctuating fields, which gives the product of determinants
\begin{align}
Z_{SU(N)}=e^{-i\theta}\left(\textbf{det} \mathcal{M}_A\right)^{-1/2}\left(\textbf{det} \mathcal{M}_{\psi}\right)\left(\textbf{det} \mathcal{M}_{\rm ghost}\right)
\left(\textbf{det} \mathcal{M}_{\phi}\right)^{-1}e^{-8\pi^2/g^2(\mu)}\, ,
\end{align}
where $\textbf{det} \mathcal{M}$ is a shorthand notation for the functional determinant of the operator $\mathcal{M}$ regulated with Pauli-Villars fields\footnote{By applying Pauli-Villars regularization, with UV regulator mass scale $\mu$, we effectively substitute the bare coupling in $\eqref{expanded action}$, with the running coupling evaluated at the UV scale $\mu$, $g(\mu)$, which satisfies the RGE
\begin{equation}
\frac{d g}{d\ln\mu_0}=-\frac{b_G^{(0)}}{16\pi^2} g^3(\mu_0)+\mathcal{O}(g^5)\, , \qquad b_G^{(0)}=\frac{11}{3}C_2[G]-\frac{2}{3}\sum_{i=1}^{\mathcal{F}}T(\textbf{R}_{i})-\frac{1}{3}\sum_{i=1}^{\mathcal{S}}T(\textbf{R}_i)\, .
\end{equation}\,}, with UV regulator mass scale $\mu$, and normalized by the zero background field determinant
\begin{equation}
\textbf{det} \mathcal{M}\equiv \frac{\det \mathcal{M} }{\det(\mathcal{M}+\mu^2)}\frac{\det(\mathcal{M}^0+\mu^2)}{\det \mathcal{M}^0}\, .
\end{equation}
These determinants can be decomposed in two parts: one corresponding to the zero modes, \textit{i.e.} the modes with zero eigenvalue of the operator $\mathcal{M}$, and the other to the non-zero modes. The expression of these determinants is well-established in the literature and a clear derivation of the non-zero modes part is given in Appendix \ref{Instanton density}.
In this work, we extend these results to matter fields in arbitrary representations of $SU(N)$, and we present it in a form that can be easily generalized to any simple gauge group.

Given that $\mathcal{M}_A$ and $\mathcal{M}_{\psi}$ have zero eigenvalues we can decompose the product of determinants as
\begin{align}
Z_{SU(N)}=& ~e^{-i\theta}\left(\text{det}_{(0)} \mathcal{M}_A\right)^{-1/2}\left(\textbf{det'} \mathcal{M}_A\right)^{-1/2}\left(\text{det}_{(0)} \mathcal{M}_{\psi}\right)\left(\textbf{det'} \mathcal{M}_{\psi}\right)\nonumber\\
\times & \left(\textbf{det} \mathcal{M}_{\rm ghost}\right)
\left(\textbf{det} \mathcal{M}_{\phi}\right)^{-1}e^{-8\pi^2/g^2(\mu)}\, ,
\end{align}
where we have split the determinants as
\begin{equation}
\textbf{det} \mathcal{M}=\left(\text{det}_{(0)} \mathcal{M}\right)\left(\frac{\text{det'} \mathcal{M} }{\text{det}(\mathcal{M}+\mu^2)}\frac{\text{det}(\mathcal{M}^0+\mu^2)}{\text{det}\mathcal{M}^0}\right)\equiv\left(\text{det}_{(0)} \mathcal{M}\right)\left(\textbf{det'} \mathcal{M}\right)\, ,
\end{equation}
where $\text{det}_{(0)} \mathcal{M}$ denotes the determinant over the zero modes only, while $\textbf{det'} \mathcal{M}$ refers to the normalized and regulated determinant over the non-zero modes.
Combining the results for the determinants over non-zero modes presented in Appendix \ref{Instanton density}, we obtain
\begin{align}
    &\left(\textbf{det'} \mathcal{M}_A\right)^{-1/2}\left(\textbf{det'} \mathcal{M}_{\psi}\right)\left(\textbf{det} \mathcal{M}_{\rm ghost}\right)
\left(\textbf{det}\mathcal{M}_{\phi}\right)^{-1}=\rho^{-4N}\Bigg(\prod_{f=1}^{\mathcal{F}}\rho^{T(\textbf{R}_{f})}\Bigg)\nonumber\\
&\times\exp\left[b_{SU(N)}^{(0)}\ln(\mu\rho)-\alpha(1)-2(N-2)\alpha(1/2)+\sum_{i\rightarrow \{\textbf{R}_{\mathcal{F}}\}}\alpha(t_i)-\sum_{i\rightarrow \{\textbf{R}_{\mathcal{S}}\}}\alpha(t_i)\right]\, .
\label{result of the determinant over non-zero modes}
\end{align}
In this expression, the first term in the exponential will promote the running coupling $g(\mu)$ in Eq. $\eqref{expanded action}$ to the running coupling evaluated at the scale $\rho^{-1}$, and the sums are taken over the isospin representations of $SU(2)$ involved in the decomposition of the generators of the representations of all scalars and fermions under the instanton corner, as explained in Appendix \ref{embedding}.

Having addressed the non-zero modes, two issues remain. The first is related to the gauge zero modes, which cause the determinant of $\mathcal{M}_A$ to vanish, leading to a divergent amplitude. The second issue arises from the fermion zero modes, which cause the amplitude to vanish.
The first problem is solved by trading the integral over the gauge zero modes for an integral over the instanton collective coordinates, which parametrize the solution in Eq. $\eqref{instanton solution}$ and can be given a clear physical interpretation. Thus, we have \cite{tHooft:1976snw,Bernard:1979qt,Vandoren:2008xg}
\begin{equation}
\left(\text{det}_{(0)} \mathcal{M}_A\right)^{-1/2}=\frac{1}{\pi^2}\frac{2^{2(1-N)}}{(N-1)!(N-2)!}\left(\frac{8\pi^2}{g^{2}}\right)^{2N}\int d^4x_0\int\frac{d\rho}{\rho^5}\rho^{4N}\int_{S^{2N-1}} d\tilde{\Omega} \, .
\end{equation}
Note that in all of the computations performed in this paper, the normalized integral over the sphere $S^{2N-1}$ will just give $1$, as in each steps of the computations we will keep the original $SU(N)$ gauge invariance. However, in the case of \textit{constrained instantons} \cite{Affleck:1980mp}, as explored in \cite{Csaki:2019vte}, this integral will have a non-trivial effect, as the vacuum-expectation-value of the scalar field that breaks $SU(N)$ depends on the instanton orientation.

The second issue means that there is no instanton contribution to the vacuum energy in a free theory with fermions. However, fermion masses and interactions lead to a non-zero result. In the following, we introduce sources for the scalar and fermion fields and shift our focus to the generating functional of the free theory, instead of the vacuum-to-vacuum amplitude. Functional derivatives with respect to these sources then allow us to systematically account for interactions.

In the background of an instanton, fermion sources are introduced as follows
\begin{align}
\rho^{T(\textbf{R})}\exp\left[-\frac{2}{3}T(\textbf{R})\ln(\mu\rho)+\sum_{i\rightarrow \textbf{R}}\alpha(t_i) \right]\Bigg(\prod_{i=1}^{2T(\textbf{R})} \frac{d\bar{a}_i}{\sqrt{\bar{\upupsilon}_{0i}}}\Bigg)\exp\left[-\int_x J^{\dagger}(x)\cdot \uppsi_0^{\dagger}\left(x\right) \right]\, ,
\end{align}
for a fermion in the representation $\textbf{R}$ of $SU(N)$, with $2T(\textbf{R})$ zero modes encapsulated in $\uppsi_0^{\dagger}$, corresponding to the set of Grassmann collective coordinates $\{\bar{a}_i\}$ with norms $\left\{\sqrt{\bar{\upupsilon}_{0i}}\right\}$, as discussed in Appendix \ref{fermion zero modes and sources}. In addition to this term, we should include a factor corresponding to the Green's function of the fermion operator, excluding the fermion zero modes. However, as explained in Appendix \ref{fermion zero modes and sources}, this factor does not contribute to our calculations, as we ultimately set $J=J^{\dagger}=0$. From now on, we will omit this exponential factor, retaining only the $J^{\dagger}$ dependence in the fermionic part of the generating functional.

Regrouping all the terms, the free generating functional of the theory in the background of an instanton is
\begin{align}
Z_0[\{J\}]=&~e^{-i\theta}\mathcal{K}_{\alpha}\int_{S^{2N-1}}d\widetilde{\Omega}\int d^4 x_0 \int\frac{d\rho}{\rho^5}\updelta_N(\rho)\prod_{f=1}^{\mathcal{F}}\Bigg\{\rho^{T(\textbf{R}_{f})}\Bigg(\prod_{i=1}^{2T(\textbf{R}_{f})}\int \frac{d \bar{a}_i}{\sqrt{\bar{\upupsilon}_{0i}}} \Bigg)\Bigg.\nonumber\\
\times & \Bigg. \exp\left[-\int_x J_{f}^{\dagger}(x)\cdot \uppsi^{\dagger}_{0,f}(x) \right]  \Bigg\} \prod_{s=1}^{\mathcal{S}}\exp\left[-\int_{x,y}J^{\dagger}_{s}(x)\cdot D_{s}(x,y)\cdot J_{s}(y) \right]\, ,
\label{final generating functional}
\end{align}
for a theory whose matter content consists of $\mathcal{F}$ Weyl fermions and $\mathcal{S}$ complex scalars in any representation of $SU(N)$, and where we introduced the coefficient
\begin{equation}
\mathcal{K}_{\alpha}=\exp\left[\sum_{i\rightarrow \{{\textbf{R}}_{\mathcal{F}}\}}\alpha(t_i)-\sum_{i\rightarrow {\{\textbf{R}}_{\mathcal{S}}\}}\alpha(t_i) \right]-\alpha(1)-2(N-2)\alpha(1/2)\, ,
\end{equation}
where $\alpha(t)$ is defined in Appendix \ref{Instanton density}. We introduced the \textit{reduced} instanton density, which is defined by extracting the $\alpha$-terms associated with the gauge bosons, differing from the usual definition of the instanton density. This is given by
\begin{equation}
\updelta_N(\rho)=\frac{1}{\pi^2}\frac{2^{2(1-N)}}{(N-1)!(N-2)!}\left(\frac{8\pi^2}{g^2}\right)^{2N}e^{-\frac{8\pi^2}{g^2(1/\rho)}}\, .
\end{equation}
From Eq. $\eqref{final generating functional}$, we observe that in the absence of interactions, any vacuum-to-vacuum amplitude in the background of an instanton vanishes due to the presence of fermions. However, this is not the final conclusion; interactions play a crucial role in saturating the integration over the Grassmann collective coordinates associated with the fermion zero modes, as we will demonstrate in the next section.

\subsection{Vacuum-to-vacuum amplitude in an interacting theory}

In the functional framework we are working with, the one-instanton generating functional in interacting theories described by $\mathcal{L}_{\rm int}$ is obtained as follows
\begin{equation}
    Z_{\rm int}[\{J\}]=\exp\left[-\int d^4x \mathcal{L}_{\rm int}\left(\left\{ -\frac{\delta}{\delta J(x)} \right\}\right) \right]Z_0[\{J\}]\, .
\end{equation}
The vacuum-to-vacuum amplitude in the interacting theory is thus given by the perturbative expansion of
\begin{equation}
    \mathcal{Z}_{SU(N)}=\Bigg.\exp\left[-\int d^4x \mathcal{L}_{\rm int}\left(\left\{ -\frac{\delta}{\delta J(x)} \right\}\right) \right]Z_0[\{J\}]\Bigg|_{\{J\}=0}\, .
    \label{vacuum to vacuum amplitude in interacting theory}
\end{equation}
Therefore, it follows that applying multiple functional derivatives with respect to sources associated with fields in the interactions makes the fermion zero modes crucial for evaluating vacuum-to-vacuum amplitudes. In the next section, we will construct these modes for any representations of $SU(2)$ and display a method to compute them for any representation of $SU(N)$ using specific examples.

%%%%%%%%%%%%%%%%%%%%%%%%%%%%%%%%%%%%%%%%%%%%%%%%%%%%%%%
\section{Fermion zero modes} \label{Fermion zero modes}

In the presence of an instanton, left-handed fermions have no zero modes, while right-handed fermions do. The reverse is true for anti-instantons. In this section, we construct all fermion zero modes for the isospin-$t$ representation of $SU(2)$ in the background of an instanton in the regular gauge and subsequently extend these results to $SU(N)$ representations. The number of these zero modes is determined by the Adler-Bell-Jackiw (ABJ) anomaly \cite{Adler:1969gk,Bell:1969ts} combined with Eq. $\eqref{instanton winding number}$. For a Weyl fermion $\psi_{\mathbf{R}}^{\textcolor{white}{\dagger}}$ in the representation $\mathbf{R}$ of $SU(N)$, and its conjugate $\psi_{\mathbf{R}}^{\dagger}$ in the representation $\mathbf{\overline{R}}$, the difference in their zero modes is given by \cite{terning2006modern}
\begin{equation}
n_{\psi^{\dagger}_{\mathbf{R}}}-n_{\psi_{\mathbf{R}}^{\textcolor{white}{\dagger}}}=2T(\mathbf{R})\frac{1}{16\pi^2}\int d^4x\text{Tr}\left[F_{\mu\nu}\widetilde{F}_{\mu\nu}\right]=2T(\mathbf{R})n\, ,
\label{index theorem}
\end{equation}
where $T(\mathbf{R})$ is the Dynkin index of the representation $\mathbf{R}$, normalized such that for the fundamental representation $T(\textbf{Fund})=1/2$, and $n$ is the Pontryagin number. 

\subsection{Fermion zero modes for isospin-$t$ representation of $SU(2)$}\label{fermion zero mode isospin-t representation}

In the background of an instanton, the fermion zero modes are defined as the set of normalizable solutions of the equation for the fermion quantum fluctuations
\begin{equation}
(\sigma_{\mu})_{\alpha\dot{\alpha}}D_{\mu}\xi^{\dot{\alpha}}=0\, ,
\end{equation}
where $D_{\mu}$ denotes the covariant derivative in the instanton background, which depends on the isospin representation of $\xi^{\dot{\alpha}}$. Given that $\xi^{\dot{\alpha}}$ is a zero mode of $\sigma_{\mu}D_{\mu}$, it is also a zero mode of $\bar{\sigma}_{\mu}\sigma_{\nu}D_{\mu}D_{\nu}$, \textit{i.e.} $\text{Ker}\left(\sigma_{\mu}D_{\mu} \right)\subset \text{Ker}\left(\bar{\sigma}_{\mu}\sigma_{\nu}D_{\mu}D_{\nu}\right)$. Consequently, we can solve the simpler equation
\begin{equation}
\left(\bar{\sigma}_{\mu}\right)^{\dot{\alpha}\beta}\left(\sigma_{\nu}\right)_{\beta\dot{\beta}}D_{\mu}D_{\nu}\xi^{\dot{\beta}}=0\, .
\end{equation}
Using Eq. $\eqref{sigmabarsigma}$ along with the instanton solution in Eq. $\eqref{instanton solution}$, we can rewrite this equation as
\begin{equation}
-\left(D^2\xi^{\dot{\alpha}}\right)^{i_1\cdots i_{2t}}+\frac{16\rho^2}{(x^2+\rho^2)^2}\left(S^{a}\right)^{\dot{\alpha}}_{\,\,\, \dot{\beta}}\left(T^{a}\xi^{\dot{\beta}}\right)^{i_1\cdots i_{2t}}=0\, ,
\label{equation fermion zero modes}
\end{equation}
where the first term is given by
\begin{equation}
\left(D^2\xi^{\dot{\alpha}}\right)^{i_1\cdots i_{2t}}=\partial^2\xi^{\dot{\alpha}i_1\cdots i_{2t}}-\frac{8}{x^2+\rho^2}L_1^{a}\left(T^{a}\xi^{\dot{\alpha}}\right)^{i_1\cdots i_{2t}}-\frac{4x^2}{(x^2+\rho^2)^2}\left(T^2\xi^{\dot{\alpha}}\right)^{i_1\cdots i_{2t}}\, ,
\label{Daucarré}
\end{equation}
and where we represent isospin-$t$ fermions as totally symmetric rank $2t$ tensors with components $\xi^{i_1\cdots i_{2t}}$. Following 't Hooft, we introduced in Eqs. $\eqref{equation fermion zero modes}$ and $\eqref{Daucarré}$ the relevant angular momenta of the problem: the spin angular momentum $\vec{S}$, represented by $\left(S^{a}\right)^{\dot{\alpha}}_{\,\,\,\dot{\beta}}$; the angular momentum $\vec{L}_1$, corresponding to the first factor of the isomorphism $SO(4)\simeq SU(2)_1\times SU(2)_2$, as detailed in Appendix \ref{Angular momenta of the problem}; the isospin angular momentum $\vec{T}$, represented by $\left(T^{a}\right)^{i}_{\,\,\, j}$; and the total angular momentum of the problem $\vec{J}_{\rm tot}=\vec{L}_1+\vec{S}+\vec{T}$. We will search for rotationaly invariant solutions of these equations, \textit{i.e.} corresponding to a zero eigenvalue of $J^2_{\rm tot}$. This means that the highest weight $\ell_1$ of $\vec{L}_1$, should be such that $\ell_1\in \llbracket 0, t-s \rrbracket$, where $t$ and $s$ are respectively the highest weights of $\vec{T}$ and $\vec{S}$.
\newline

The sign of the eigenvalues of the coupling $\vec{S} \cdot \vec{T}$ indicates whether the equation admits solutions. Specifically, if we consider a tensor associated with a positive eigenvalue of $\vec{S} \cdot  \vec{T}$, Eq. $\eqref{equation fermion zero modes}$ simplifies to a sum of positive definite operators, which consequently lack zero modes \cite{Vandoren:2008xg,Shifman:2012zz,Tong:2005un}. The presence of negative eigenvalues enables us, as we will see, to construct fermion zero modes. The eigenvalue equation for the $\vec{S} \cdot  \vec{T}$ operator is solved by the tensor
\begin{equation}
\varphi^{\dot{\alpha}_1i_1\cdots i_{2t}}=\mathcal{A}^{\dot{\alpha}_1\cdots \dot{\alpha}_{2t},i_1\cdots i_{2t}}\mathcal{M}_{\dot{\alpha}_2\cdots \dot{\alpha}_{2t}}\, ,\quad\text{with}\quad \mathcal{A}^{\dot{\alpha}_1\cdots \dot{\alpha}_{2t},i_1\cdots i_{2t}}=\frac{1}{(2t)!}\sum_{\sigma\in\mathfrak{S}_{2t}}\prod_{k=1}^{2t}\varepsilon^{i_{\sigma(k)}\alpha_k}\, ,
\end{equation}
where $\mathfrak{S}_{2t}$ denotes the rank $2t$ group of permutations and $\mathcal{M}$ is a rank $2t-1$ totally symmetric tensor of Grassmann coefficients. Since $\vec{S}\cdot \vec{T}$ commutes with $T^2$, this tensor also solves the eigenvalue equation of $T^2$ and we have
\begin{equation}
\left(S^{a}\right)^{\dot{\alpha}}_{\,\,\,\dot{\beta}}\left(T^{a}\varphi^{\dot{\beta}}\right)^{i_1\cdots i_{2t}}=-\frac{1}{2}(t+1)\varphi^{\dot{\alpha}i_1\cdots i_{2t}},\quad\text{and}\quad \left(T^{2}\varphi^{\dot{\alpha}}\right)^{i_1\cdots i_{2t}}=t(t+1)\varphi^{\dot{\alpha}i_1\cdots i_{2t}}\, .
\end{equation}
The remaining operators to diagonalize are $\vec{L}_1\cdot \vec{T}$ and $L_1^2$, the latter of which appears in the $4d$ Laplace operator. They are diagonalized by introducing certain representations of the $4d$ spherical harmonics in Cartesian coordinates, specifically tensor products of $(x\cdot\bar{\sigma})$, as discussed in Appendix \ref{Angular momenta of the problem}. The eigenvalue equations for the operators $\vec{L}_1\cdot \vec{T}$ and $L_1^2$ are solved by the tensor
\begin{equation}
\upphi^{\dot{\alpha}_1 i_1\cdots i_{2t}}_{(\ell_1)}(x)=f_{2t}(r)\mathcal{A}^{\dot{\alpha}_1\cdots \dot{\alpha}_{2t},i_1\cdots i_{2t}}\left(\prod_{i=2}^{2\ell_1+1}(x\cdot\bar{\sigma})_{\dot{\alpha}_i \beta_i}\right)\left(\prod_{j=2\ell_1+2}^{2t}\delta_{\dot{\alpha}_j}^{\dot{\beta}_j}\right)\mathcal{M}^{(2\ell_1)\,\, \beta_2\cdots \beta_{2\ell_1+1}}_{\textcolor{white}{(2\ell_1)\,\,}\dot{\beta}_{2\ell_1+2}\cdots \dot{\beta}_{2t}}\, .
\label{solution l1 and t}
\end{equation}
In this expression, $f_{2t}(r)$ is an isospin-dependent function that encapsulates the radial dependence of the solution, and $\mathcal{M}^{(k)\,\, \beta_2\cdots \beta_{k+1}}_{\textcolor{white}{(k)\,\,}\dot{\beta}_{k+2}\cdots \dot{\beta}_{2t}}$ is a rank $2t-1$ tensor of Grassmann variables that is fully symmetric under permutations of its $k$ undotted indices and likewise for its $2t-k-1$ dotted indices. The eigenvalue equations are of the form
\begin{equation}
L_1^{a}(T^{a}\upphi_{(\ell_1)}^{\dot{\alpha_1}})^{i_1\cdots i_{2t}}=-\ell_1(t+1)\upphi_{(\ell_1)}^{\dot{\alpha_1}i_1\cdots i_{2t}},\quad\text{and}\quad L_1^2\upphi_{(\ell_1)}^{\dot{\alpha_1}i_1\cdots i_{2t}}=\ell_1(\ell_1+1)\upphi_{(\ell_1)}^{\dot{\alpha_1}i_1\cdots i_{2t}}\, .
\end{equation}
Therefore, the equation of motion for this tensor becomes
\begin{equation}
\frac{\partial^{2}\upphi_{(\ell_1)}^{\dot{\alpha}}}{\partial r^2}+\frac{3}{r}\frac{\partial \upphi_{(\ell_1)}^{\dot{\alpha}}}{\partial r}-\frac{4}{r^2}\ell_1(\ell_1+1)\upphi_{(\ell_1)}^{\dot{\alpha}}+\frac{8\ell_1(t+1)}{r^2+\rho^2}\upphi_{(\ell_1)}^{\dot{\alpha}}-\frac{4t(t+1)r^2}{(r^2+\rho^2)^2}\upphi_{(\ell_1)}^{\dot{\alpha}}+\frac{8(t+1)\rho^2}{(r^2+\rho^2)^2}\upphi_{(\ell_1)}^{\dot{\alpha}}=0\, ,
\end{equation}
which is solved for
\begin{equation}
f_{2t}(r)=\frac{1}{(r^2+\rho^2)^{t+1}}\, .
\label{solution f(r)}
\end{equation}
The complete form of the fermion zero modes that are eigenfunctions of all the operators of the problem and corresponding to $j_{\rm tot}=0$, which requires that $\ell_1\in \llbracket 0, t-1/2\rrbracket$, can be written as
\begin{equation}
\xi^{\dot{\alpha}_1i_1\cdots i_{2t}}(x)=f_{2t}(r)\mathcal{A}^{\dot{\alpha}_1\cdots \dot{\alpha}_{2t},i_1\cdots i_{2t}}\sum_{k=0}^{2t-1}\left(\prod_{i=2}^{2\ell_1+1}(x\cdot\bar{\sigma})_{\dot{\alpha}_i \beta_i}\right)\left(\prod_{j=2\ell_1+2}^{2t}\delta_{\dot{\alpha}_j}^{\dot{\beta}_j}\right)\mathcal{M}^{(2\ell_1)\,\, \beta_2\cdots\beta_{2\ell_1+1}}_{\textcolor{white}{(2\ell_1)\,\,}\dot{\beta}_{2\ell_1+2}\cdots \dot{\beta}_{2t}}\, .
\label{first version of fermion zero modes isospin t}
\end{equation}
where $\mathcal{M}^{(2\ell_1)}$ contains the Grassmann collective coordinates associated to the fermion zero mode.
\newline

\noindent\textit{Normalization}
\newline

We normalize the zero modes using the following norm, which applies when the fermion’s kinetic term is canonically normalized
\begin{equation}
    \left\langle \upphi_{(\ell_1)}\left| \upphi_{(\ell_1)}\right.\right\rangle =\int d^4x\, \varepsilon_{\dot{\alpha}_1\dot{\beta}_1}\varepsilon_{i_1 j_1}\cdots\varepsilon_{i_{2t}j_{2t}}\upphi_{(\ell_1)}^{\dot{\alpha}_1i_1\cdots i_{2t}}(x)\upphi_{(\ell_1)}^{\dot{\beta}_1j_1\cdots j_{2t}}(x)\, .
    \label{norm 1}
\end{equation}
Using Eq. $\eqref{solution l1 and t}$ we obtain
\begin{align}
    \left\langle \upphi_{(\ell_1)}\left| \upphi_{(\ell_1)}\right.\right\rangle =\frac{2t+1}{2t}\frac{2\pi^2}{\rho^{4(t-\ell_1)}}\frac{\Gamma(2\ell_1+2)\Gamma(2t-2\ell_1)}{2\Gamma(2t+2)}\mathcal{A}^{\dot{\alpha}_2\cdots\dot{\alpha}_{2t},\dot{\beta}_2\cdots\dot{\beta}_{2t}}\mathcal{M}^{(2\ell_1)}_{\dot{\alpha}_2\cdots\dot{\alpha}_{2t}}\mathcal{M}^{(2\ell_1)}_{\dot{\beta}_2\cdots\dot{\beta}_{2t}}\, .
\end{align}
Thus, we observe that all Grassmann collective coordinates corresponding to a given $\ell_1$, contained in the tensor $\mathcal{M}^{(2\ell_1)}$, share the same norm, which is expressed as
\begin{equation}
   \upupsilon_{\upphi_{(\ell_1)}}\equiv\frac{2t+1}{2t}\frac{2\pi^2}{\rho^{4(t-\ell_1)}}\frac{\Gamma(2\ell_1+2)\Gamma(2t-2\ell_1)}{2\Gamma(2t+2)}\, ,
    \label{norm 2}
\end{equation}
to match the expression given in Eq. $\eqref{original definition of the norms}$.
In addition, two isospin-$t$ fermion zero modes associated to different eigenvalues of $L_1^2$ are orthogonal, in other words
\begin{equation}
     \left\langle \upphi_{(\ell_1)}\left| \upphi_{(\ell_2)}\right.\right\rangle = 0\qquad\text{for}\qquad \ell_1\ne \ell_2\, .
\end{equation}
\noindent \textit{Number of zero modes}
\newline

We can now check that we have found the right number of zero modes.
The dimension of the space\footnote{The dimension of the space of all fully symmetric tensors of rank $k$ defined on a vector space of dimension $N$ is given by the binomial coefficient $\binom{N+k-1}{k}$. For $N=2$ we simply have $k+1$.} of the tensors $\mathcal{M}_{\textcolor{white}{(k)\,\,}\dot{\alpha}_{k+2}\cdots \dot{\alpha}_{2t}}^{(k)\,\,\alpha_2\cdots \alpha_{k+1}}$ is $(k+1)\times (2t-k)$, thus the number of fermion zero modes contained in Eq. $\eqref{first version of fermion zero modes isospin t}$ is
\begin{equation}
\sum_{k=0}^{2t-1}(k+1)(2t-k)=\frac{2}{3}t(t+1)(2t+1)=2T(t)\, ,
\end{equation}
which is nothing but twice the Dynkin index of the isospin representation $t$, as expected from Eq. $\eqref{index theorem}$. However, this result has been derived for a specific orientation of the instanton solution in $SU(2)$, which is also centered at $x_0=0$. In this configuration, the gauge field takes the form
\begin{equation}
    A_{\mu}^{SU(2)}(x)=2\eta_{a\mu\nu}T^{a}\frac{x_{\nu}}{x^2+\rho^2}\, .
\end{equation}
In the general case, the expression for the fermion zero modes in the isospin-$t$ representation is modified to account for an arbitrary instanton orientation and center position, resulting in
\begin{equation}
\uppsi^{\dot{\alpha}_1i_1\cdots i_{2t}}(x)=U^{i_1}_{\,\,\,\, j_1}(\vec\theta)\cdots U^{i_{2t}}_{\,\,\,\, j_{2t}}(\vec\theta)\xi^{\dot{\alpha}_1j_1\cdots j_{2t}}(x-x_0)\, , \qquad U\in SU(2)\, .
\label{final fermion zero mode}
\end{equation}
Thus, we have obtained the complete set of isospin-$t$ fermion zero modes in the background of the rotated instanton given in Eq. \eqref{instanton solution}, corresponding to a rotationally invariant solution. In the computation of vacuum-to-vacuum amplitudes, the $U$'s generally do not contribute, as only gauge-invariant operators are involved.
\newline

\noindent \textbf{Example:} Isospin-$1$ representation and Super(conformal)symmetry
\newline

In this example we will see that instantons share an intimate relationship with supersymmetry.
The form of the fermion zero modes in the case of the isospin-$1$ representation is well known in the litterature \cite{Novikov:1983ek,Vandoren:2008xg}. From our analysis we obtain
\begin{equation}
\upphi^{\dot{\alpha}_1i_1 i_2}_{(0)}=f_2(r)\mathcal{A}^{\dot{\alpha}_1\dot{\alpha}_2,i_1 i_2}\mathcal{M}^{(0)}_{\dot{\alpha}_2}=\frac{1}{2}f_2(r)\left(\varepsilon^{i_1\dot{\alpha}_1}\varepsilon^{i_2\dot{\alpha}_2}+\varepsilon^{i_2\dot{\alpha}_1}\varepsilon^{i_1\dot{\alpha}_2}\right)\mathcal{M}^{(0)}_{\dot{\alpha}_2}\, .
\end{equation}
However, from Eq. $\eqref{sigmabarmunusigmabarmunu}$ and using the explicit expression of $f_2(r)$ we see that
\begin{equation}
\upphi^{\dot{\alpha}_1i_1 i_2}_{(0)}=-\frac{1}{2}\frac{1}{(r^2+\rho^2)^2}\left(\bar{\sigma}_{\mu\nu}\right)^{\dot{\alpha}_1\dot{\alpha}_2}\left(\bar{\sigma}_{\mu\nu}\right)^{i_1 i_2}\mathcal{M}^{(0)}_{\dot{\alpha}_2}\propto\left(\bar{\sigma}_{\mu\nu}\right)^{\dot{\alpha}_1\dot{\alpha}_2}\left(F_{\mu\nu}\right)^{i_1 i_2}\mathcal{M}^{(0)}_{\dot{\alpha}_2}\, ,
\end{equation}
where we introduced the field strength of the instanton solution. This corresponds to an on-shell supersymmetric relation between a fermion in the adjoint representation and the instanton solution. Moreover, we also have
\begin{equation}
\upphi^{\dot{\alpha}_1i_1 i_2}_{(1/2)}=f_2(r)\mathcal{A}^{\dot{\alpha}_1\dot{\alpha}_2,i_1 i_2}(x\cdot \bar{\sigma})_{\dot{\alpha}_2 \beta_2}\mathcal{M}^{(1)\,\beta_2}\, ,
\end{equation}
which is nothing but a superconformal transformation relation between a fermion in the adjoint representation and the instanton solution
\begin{equation}
\upphi^{\dot{\alpha}_1i_1 i_2}_{(1/2)}\propto\left(\bar{\sigma}_{\mu\nu}\right)^{\dot{\alpha}_1\dot{\alpha}_2}(x\cdot \bar{\sigma})_{\dot{\alpha}_2 \beta_2}\left(F_{\mu\nu}\right)^{i_1 i_2}\mathcal{M}^{(1)\,\beta_2}\, .
\end{equation}
It is as if we had rediscovered (half of) the supersymmetry and superconformal transformations as accidental symmetries of the theory. The other half of these symmetry transformations annihilate the instanton solution \cite{Novikov:1983ek,Tong:2005un}.

\subsection{Fermion zero modes for any representation of $SU(N)$}

\begin{figure}[!htb]
\begin{center}
\includegraphics[width=\textwidth]{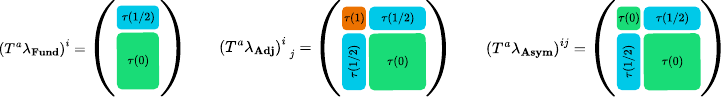}
\caption{Decomposition of the action of the $T^{a}$'s on the fermion zero modes in terms of $SU(2)$ irreps $\tau(t_i)$ for the fundamental, adjoint and antisymmetric representations of $SU(N)$.}
\label{decomposition instanton corner}
\end{center}
\end{figure}

In this section, we extend the previous results to fermions in representations of $SU(N)$. Rather than deriving the explicit form of the fermion zero modes for general representations, we focus on formulating the equation of motion and outlining the general strategy to solve it. We then apply this approach to specific representations, demonstrating how the method can be straightforwardly generalized to higher representations of $SU(N)$. We once again work in the background of an instanton, where the equation of motion for the right-handed zero modes $\uplambda^{\dot{\alpha}}$ in a representation $\textbf{R}$ of $SU(N)$, carrying $n$ upper indices and $m$ lower indices, is
\begin{equation}
\left(\sigma_{\mu}\right)^{\dot{\alpha}}_{\,\,\, \dot{\beta}}\left(D_{\mu}\uplambda^{\dot{\beta}}\right)^{i_1\cdots i_{n}}_{j_1\cdots j_m}=0\, .
\end{equation}
As in the case of the $SU(2)$ fermion zero modes, we instead solve
\begin{equation}
-\left(D^2\uplambda^{\dot{\alpha}}\right)^{i_1\cdots i_{n}}_{j_1\cdots j_m}+\frac{16\rho^2}{(x^2+\rho^2)^2}\left(S^{a}\right)^{\dot{\alpha}}_{\,\,\,\dot{\beta}}\left(T^{a}\uplambda^{\dot{\beta}}\right)^{i_1\cdots i_{n}}_{j_1\cdots j_m}=0\, ,
\end{equation}
where the expression of $D^2\uplambda$ mirrors that in Eq. $\eqref{Daucarré}$. We solve this equation within the minimal embedding framework, where the instanton is placed in the upper-left corner of the fundamental representation of $SU(N)$. As a result, the equation of motion for the zero modes decomposes into a block form, reflecting the decomposition of the fermion representation under the $SU(2)$ instanton corner
\begin{equation}
    T^{a}(\textbf{R})=\bigoplus_{i\rightarrow \textbf{R}}\tau^{a}(t_i)\, ,
\end{equation}
as illustrated in Figure \ref{decomposition instanton corner}.
Consequently, the problem of finding fermion zero modes in representations of $SU(N)$ reduces to to solving for the zero modes in $SU(2)$ representations. As in the $SU(2)$ case, the $\vec{S}\cdot\vec{T}$ coupling is essential in determining the solutions, as only tensors with negative eigenvalues lead to a zero mode.
%%%%%%%%%%%%%%%%%%%%%%%%%%%%
%%%%%%%%%%%%%%%%%%%%%%%%%%%%
For $SU(N)$ representations, this operator is diagonalized within each $SU(2)$ irreps of the block decomposition, using the results derived in Section \ref{fermion zero mode isospin-t representation}. The final result is then reconstructed by restoring the symmetry properties of the indices of the original fermion zero mode representation, based on the eigentensors of the $\vec{S}\cdot \vec{T}$ coupling. The remaining operators in $-D^2$, namely $\vec{L}_1\cdot\vec{T}$ and $L_1^2$, are diagonalized in a similar manner as in Section \ref{fermion zero mode isospin-t representation}, by contracting factors of Cartesian spherical harmonics $\left(x\cdot\bar{\sigma}\right)$ with the eigentensors of the $\vec{S}\cdot\vec{T}$ coupling.

\subsubsection{The fundamental representation}

Under the instanton corner, the fundamental representation of $SU(N)$ decomposes as
\begin{equation}
    T^{a}(\textbf{Fund})=\tau^{a}(1/2)\oplus (N-2)\tau^{a}(0)\, ,
    \label{decomposition of fundamental of SU(N)}
\end{equation}
resulting in the equation of motion for the zero modes splitting into independent equations, each corresponding to a term in this direct sum. As illustrated in Figure \ref{decomposition instanton corner}, the action of the $T^{a}$'s in the equation of motion eliminates all the components of $\uplambda^{i}$ except those associated with the $\tau^{a}(1/2)$ term in Eq. $\eqref{decomposition of fundamental of SU(N)}$. Therefore, only this component will give rise to a zero mode, as the remaining components involve only positive definite operators. The diagonalization of the $\vec{S}\cdot\vec{T}$ coupling is then carried out in the same manner as for the fundamental representation of $SU(2)$, and we obtain
\begin{equation}
    \uplambda^{\dot{\alpha}i}(x)=f_{1}(r)\varepsilon^{\dot{\alpha} i}\mathcal{M}^{(0)}\, ,
\end{equation}
where $\mathcal{M}^{(0)}$ is a Grassmann collective coordinate associated to the fermion zero mode. The expression for the anti-fundamental representation is
\begin{equation}
    \uplambda^{\dot{\alpha}}_i(x)=f_{1}(r)\delta^{\dot{\alpha}}_i\mathcal{M}^{(0)}\, .
\end{equation}
To write these embedded solutions we introduced the embedded Kronecker-delta and Levi-Civita symbol
\begin{equation}
\delta^{\dot{\alpha}}_i=
\begin{pmatrix}
 && 0 && \cdots && 0\\
\delta^{\dot{\alpha}}_{\dot{\beta}} &&  && && \\
 && 0 && \cdots && 0\\
\end{pmatrix} , \qquad \varepsilon^{\dot{\alpha}i}=
\begin{pmatrix}
 && 0 && \cdots && 0\\
\varepsilon^{\dot{\alpha}\dot{\beta}} &&  && && \\
 && 0 && \cdots && 0\\
\end{pmatrix},\qquad \dot{\alpha},\dot{\beta}\in \llbracket 1,2\rrbracket\, .
\end{equation}
where $\delta^{\dot{\alpha}}_{\dot{\beta}}$ and $\varepsilon^{\dot{\alpha}\dot{\beta}}$ are the usual two-dimensional symbols.

\subsubsection{The adjoint representation}

Under the instanton corner, the adjoint representation of $SU(N)$ decomposes as
\begin{equation}
    T^{a}(\textbf{Adj})=\tau^{a}(1)\oplus 2(N-2)\tau^{a}(1/2)\oplus (N-2)^2\tau^{a}(0)\, .
\end{equation}
From Figure \ref{decomposition instanton corner}, we observe that this decomposition leads to one equation of motion corresponding to the isospin-$1$ representation of $SU(2)$, along with $2(N-2)$ equations associated with the isospin-$1/2$ representation. The remaining components, which involve positive definite operators, do not yield zero modes. The adjoint representation has $2T(\textbf{Adj})=2N$ zero modes, consisting of four from the isospin-$1$ component and $2N-4$ from the isospin-$1/2$ sector. The equation for the isospin-$1$ yields a solution localized at the instanton corner, given by
\begin{equation}
    \left(\uplambda_{\textbf{Adj}}^{\dot{\alpha}_1}\right)^{i_1}_{\,\,\, i_2}(x)=f_2(r)\mathcal{A}^{\dot{\alpha}_1\dot{\alpha}_2,i_1}_{\textcolor{white}{\dot{\alpha}_1\dot{\alpha}_2,i_1}i_2}\left(\mathcal{M}_{\dot{\alpha}_2}^{(0)}+\left(x\cdot\bar{\sigma}\right)_{\dot{\alpha}_2}^{\,\,\,\beta_2}\mathcal{M}_{\beta_2}^{(1)} \right)\, ,\qquad i_1,i_2\in\llbracket 1,2\rrbracket \, .
\end{equation}
The other solutions correspond to the fundamental and anti-fundamental representations of $SU(2)$, they have the form
\begin{equation}
    \left(\uplambda_{\textbf{Adj}}^{\dot{\alpha}}\right)^{i_1}_{\,\,\, i_2}(x)=f_1(r)\left(\upomega^{i_1}\delta^{\dot{\alpha}}_{i_2}+\varepsilon^{\dot{\alpha}i_1}\bar{\upomega}_{i_2}\right)\, ,
\end{equation}
where the Grassmann collective coordinates are encapsulated within the vectors
\begin{equation}
    \upomega^{i}=
    \begin{pmatrix}
        0 && 0 && \upomega^3 && \cdots && \upomega^{N}\\
    \end{pmatrix},\qquad \bar{\upomega}_{i}=
    \begin{pmatrix}
        0 && 0 && \bar{\upomega}_3 && \cdots && \bar{\upomega}_{N}\\
    \end{pmatrix}\, .
    \label{Grassmann vectors adjoint}
\end{equation}

\subsubsection{The antisymmetric representation}

To compute instanton contributions in models such as the minimal $SU(5)$ GUT, we need the expression for the fermion zero modes in the antisymmetric representation of $SU(5)$, which contains some quarks and leptons of the Standard Model. Under the $SU(2)$ instanton corner, the antisymmetric representation of $SU(N)$ decomposes\footnote{This follows from the decomposition $\textbf{N}\otimes \textbf{N}=\textbf{Sym}\oplus\textbf{Asym}$ and dimension counting, given that $\textbf{Asym}$ has dimension $N(N-1)/2$ and $\textbf{Sym}$ has dimension $N(N+1)/2$.} as
\begin{equation} 
T^{a}(\textbf{Asym}) = (N-2)\tau^{a}(1/2) \oplus \left(\frac{N^2-5N+8}{2}\right)\tau^{a}(0) \, .
\label{direct sum asym} 
\end{equation} 
From Figure \ref{decomposition instanton corner}, we see that this decomposition results in $N-2$ equations of motion corresponding to the isospin-$1/2$ representation of $SU(2)$. The remaining components, involving positive definite operators, do not produce zero modes. The antisymmetric representation of $SU(N)$ contains $2T(\textbf{Asym}) = N-2$ zero modes, all of which arise from the $N-2$ copies of the isospin-$1/2$ representation of $SU(2)$. The zero modes for the antisymmetric representation are constructed by restoring antisymmetry among the gauge group indices, using the $N-2$ copies of the isospin-$1/2$ eigentensor of the $\vec{S}\cdot \vec{T}$ coupling as building blocks, which leads to
\begin{equation}
\uplambda^{\dot{\alpha}i_1 i_2}=f_1(r)\left[\varepsilon^{\dot{\alpha}i_1}\upmu^{i_2}-\varepsilon^{\dot{\alpha}i_2}\upmu^{i_1} \right],\qquad \upmu=
\begin{pmatrix}
0 && 0 && \upmu^{3} && \cdots && \upmu^N\\
\end{pmatrix}\, ,
\label{fermion zero mode antisymmetric}
\end{equation}
where $f_1(r)$ is given in Eq. $\eqref{solution f(r)}$ and $\upmu$ is a vector containing the Grassmann collective coordinates.

\section{Examples} \label{Examples}

In this section, we will apply the functional method outlined in the first part of the paper to evaluate instanton contributions to the axion potential in both the MSSM and the Minimal Supersymmetric $SU(5)$ GUT. We will conduct a detailed computation for $SU(2)_L$, evaluating all relevant diagrams, while our approach for supersymmetric QCD and SUSY $SU(5)$ will be more streamlined. Although supersymmetry is not a requirement for these computations, we have chosen to work within this framework to provide a more comprehensive analysis compared to the non-supersymmetric case, as the particle spectrum is richer.

\subsection{Gaugino mass as an interaction}

To illustrate the functional method, we will consider the toy example of a Supersymmetric $SU(N)$ Yang-Mills theory\footnote{In Euclidean space $\lambda$ and $\lambda^{\dagger}$ are treated as independent variables, which makes difficult to define a real action. For the purposes of this paper, we set this issue aside and refer the reader to \cite{Dorey:2002ik,Shifman:2012zz} for a detailed discussion of the problem and its solutions.}, omitting the auxiliary field and treating the SUSY-breaking gaugino mass as an interaction term. The Lagrangian that we consider in Minkowski space-time is given by
\begin{equation}
    \mathcal{L}_{\rm SYM}\supset \frac{1}{g^2}\text{Tr}\left[-\frac{1}{2}F^{\mu\nu}F_{\mu\nu}+2i\lambda^{\dagger}\bar{\sigma}^{\mu}D_{\mu}\lambda-\left(\widetilde{M}\lambda\lambda+\text{h.c.}\right)\right]\, .
\end{equation}
With this particle content, the free generating functional is given by
\begin{align}
    Z_0\left[J^{\dagger}\right]=e^{-i\theta}\int d^4x_0\int\frac{d\rho}{\rho^5}\updelta_N(\rho)\rho^{T(\textbf{Adj})}\int \frac{d^2\upxi}{\bar{\upupsilon}_{\upxi}} \frac{d^2\upeta}{\bar{\upupsilon}_{\upeta}}\prod_{u=3}^N\left(\frac{d\upomega^{u}d\bar{\upomega}_u}{\bar{\upupsilon}_{\upomega}}\right)\exp\left[-\int_x J^{\dagger}\cdot\uppsi_{\lambda}^{\dagger} \right]\, .
    \label{Generating functional for gaugino mass term}
\end{align}
Here, $\mathcal{K}_{\alpha}=1$ due to the supersymmetric nature of the model, and we have encapsulated the fermion zero modes of the adjoint representation into $\uppsi^{\dagger}_{\lambda}$, which take the following form
\begin{equation}
    \left(\lambda^{\dot{\alpha}_1}_{\upxi}\right)^{i_1}_{\,\,\, i_2}(x)=f_2(r)\mathcal{A}^{\dot{\alpha}_1\dot{\alpha}_2,i_1}_{\textcolor{white}{\dot{\alpha}_1\dot{\alpha}_2,i_1}i_2} \upxi_{\dot{\alpha}_2}\, ,\qquad \left(\lambda^{\dot{\alpha}_1}_{\upeta}\right)^{i_1}_{\,\,\, i_2}(x)=f_2(r)\mathcal{A}^{\dot{\alpha}_1\dot{\alpha}_2,i_1}_{\textcolor{white}{\dot{\alpha}_1\dot{\alpha}_2,i_1}i_2}\left(x\cdot\bar{\sigma}\right)_{\dot{\alpha}_2}^{\,\,\, \beta_2} \upeta_{\beta_2}\, ,
\end{equation}
and,
\begin{equation}
    \left( \lambda^{\dot{\alpha}_1}_{\upomega}\right)^{i_1}_{\,\,\, i_2}(x)=f_1(r)\upomega^{i_1}\delta^{\dot{\alpha}_1}_{i_2}\, , \qquad  \left( \lambda^{\dot{\alpha}_1}_{\bar{\upomega}}\right)^{i_1}_{\,\,\, i_2}(x)=f_1(r)\varepsilon^{\dot{\alpha}_1 i_1}\bar{\upomega}_{i_2}\,,
\end{equation}
where the Grassmann vectors $\upomega^{i}$ and $\bar{\upomega}_i$ are defined in Eq. $\eqref{Grassmann vectors adjoint}$. It is important to note that the fermions are not canonically normalized. To account for the normalization of the kinetic terms in the norm, we must multiply Eqs. $\eqref{norm 1}$ and $\eqref{norm 2}$ by $2/g^{2}$, which gives
\begin{equation}
  \bar{\upupsilon}_{\upxi}\bar{\upupsilon}_{\upeta} (\bar{\upupsilon}_{\upomega} \bar{\upupsilon}_{\bar{\upomega}})^{(N-2)/2}=\left(\frac{2}{g^2}\right)^{N}\left(\frac{\pi^2}{4\rho^4}\right)\left(\frac{\pi^2}{2\rho^2}\right)\left(\frac{\pi^2}{\rho^2}\right)^{N-2}\, .
\end{equation}
Considering the measure over Grassmann collective coordinates in Eq. $\eqref{Generating functional for gaugino mass term}$, we see that we need to expand the exponential in $\mathcal{Z}_{SU(N)}$ up to oder $N$ in the gaugino mass to yield a non-zero result. This gives
\begin{align}
    \mathcal{Z}_{SU(N)}=&e^{-i\theta}\int d^4x_0\int \frac{d\rho}{\rho^5}\updelta_N(\rho)\rho^N\int \frac{d^2\upxi}{\bar{\upupsilon}_{\upxi}} \frac{d^2\upeta}{\bar{\upupsilon}_{\upeta}}\prod_{u=3}^N\left(\frac{d\upomega^{u}d\bar{\upomega}_u}{\bar{\upupsilon}_{\upomega}}\right)\frac{1}{N!}\left[\frac{\widetilde{M}}{2g^2}\int_x \uppsi_{\lambda}^{\dagger}\cdot \uppsi_{\lambda}^{\dagger} \right]^N\, .
\end{align}
%%%%%%%%%%%%%%%%%%%%%%%%%%%%
\begin{figure}[!t]
\begin{center}
\includegraphics[width=0.45\textwidth]{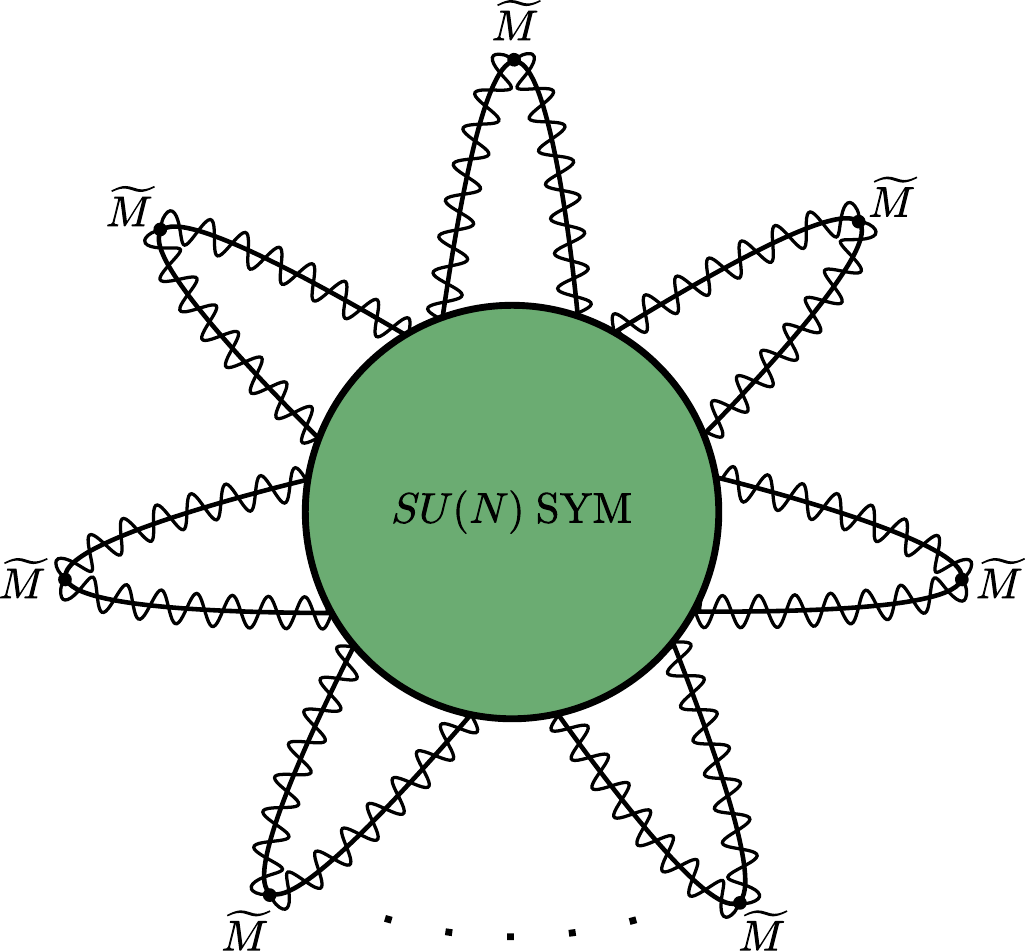}
\caption{Gaugino mass contribution to the vacuum-to-vacuum amplitude in $\mathcal{N}=1$ SYM. There are $N$ gaugino legs in the diagram from their $SU(N)$ Dynkin index.}
\label{fig:gaugino mass}
\end{center}
\end{figure}
%%%%%%%%%%%%%%%%%%%%%%%%%%%%
This contribution can be illustrated using a 't Hooft diagram, as shown in Figure \ref{fig:gaugino mass}.
Using the orthogonality property of the zero modes along with the multinomial expansion, we obtain
\begin{align}
&\int d^2\upxi d^2\upeta\prod_{u=3}^N \left(d\upomega^{u}d\bar{\upomega}_u\right)\left[\int_x \,\uppsi_{\lambda}^{\dagger}\cdot \uppsi_{\lambda}^{\dagger} \right]^N\nonumber\\
&=4\frac{N!}{(N-2)!}\prod_{u=3}^N\left(d\upomega^{u}d\bar{\upomega}_u\right)\left(\frac{\pi^2}{4\rho^4}\right)\left(\frac{\pi^2}{2\rho^2}\right) \left(\frac{2\pi^2}{\rho^2}\right)^{N-2}\left[\bar{\upomega}_v\upomega^v \right]^{N-2}\nonumber\\
&= 2^N N! \left(\frac{\pi^2}{4\rho^4}\right)\left(\frac{\pi^2}{2\rho^2}\right) \left(\frac{\pi^2}{\rho^2}\right)^{N-2}\, .
\end{align}
Combining all the results we obtain the instanton contribution to the vacuum energy from closing the gaugino legs with their masses
\begin{equation}
    \mathcal{Z}_{SU(N)}=e^{-i\theta}\int d^4 x_0 \int \frac{d\rho}{\rho^5}\updelta_N(\rho)(\rho\widetilde{M})^N\, ,
\end{equation}
which aligns with the result derived using instanton NDA \cite{Csaki:2023ziz}.

\subsection{SUSY $SU(2)_L$ + color triplets $T_{u,d}$}

It is a well-known fact that the Standard Model Electroweak $\theta$-term
\begin{equation}
    \mathcal{L}_{\theta}\supset -\frac{\theta_{\rm EW}}{16\pi^2}\text{Tr}\left[W\widetilde{W} \right]\, ,
\end{equation}
has no physical impact, as it can be eliminated through an appropriate anomalous $U(1)_{B+L}$ field redefinition of the quarks and leptons \cite{FileviezPerez:2014xju}. However, when explicit $U(1)_{B+L}$ symmetry breaking is introduced—such as by embedding the SM in a Grand Unified Theory—the parameter $\theta_{\rm EW}$ can indeed have a physical significance and contribute to the potential of axions through instantons. For this reason, we consider the Supersymmetric $SU(2)_L$ theory with two color triplets $T_{u,d}$, which arise from the spontaneous symmetry breaking of $SU(5)$. 
We will not compute the instanton-induced axion potential using a manifestly supersymmetric framework as proposed in \cite{Choi:1998ep} for supersymmetric theories. Instead, we employ the functional method introduced in the previous section, which relies on the interaction Lagrangian of the theory. Contributions to the axion potential arise from instantons of all sizes; in general to compute these contributions, we work with a series of effective theories, each valid at different energy scales. The calculations in this section are valid at energies between the SUSY-breaking scale $\widetilde{M}_S$ and the GUT scale $M_{\rm GUT}$. The superpotential we consider is\footnote{We denote superfields as their scalar component.}
\begin{equation}
    W_{SU(2)_L}\supset \frac{1}{2}Y_u \varepsilon_{abc}\varepsilon_{ij}\widetilde{Q}^{ai}\widetilde{Q}^{bj}T_u^c+Y_d \varepsilon_{ij}\widetilde{Q}^{ai}\widetilde{L}^j T_{d,a}+\mu H_u H_d\, ,
\end{equation}
where we have omitted Yukawa couplings involving the Higgs doublets, as they do not contribute to closing the fermion legs of the instanton. In addition to the superpotential, we must also consider the Yukawa-gauge interactions for each chiral supermultiplet. For a given supermultiplet $\Phi = (\phi, \psi, F)$, this interaction takes the form \cite{Martin:1997ns}
\begin{equation}
\mathcal{L}_{\rm int}\supset -\sqrt{2}\left(\phi^* \tilde{g}\,\psi+\text{h.c.}\right)\, .
\end{equation}
Along with these interactions, we also include the following soft SUSY-breaking terms
\begin{equation}
    \mathcal{L}_{\rm soft}\supset -\frac{1}{2g^2}\widetilde{M}_2 \widetilde{W}\widetilde{W}-B\mu \left(T_{u}T_d+H_u H_d\right)+\text{h.c.}\, ,
\end{equation}
as they are required to have a non-zero vacuum energy \cite{Choi:1998ep,Dine:1986bg}.
\newline

\begin{figure}[!t]
\begin{center}
     \begin{subfigure}[b]{0.35\textwidth}
         \includegraphics[width=\textwidth]{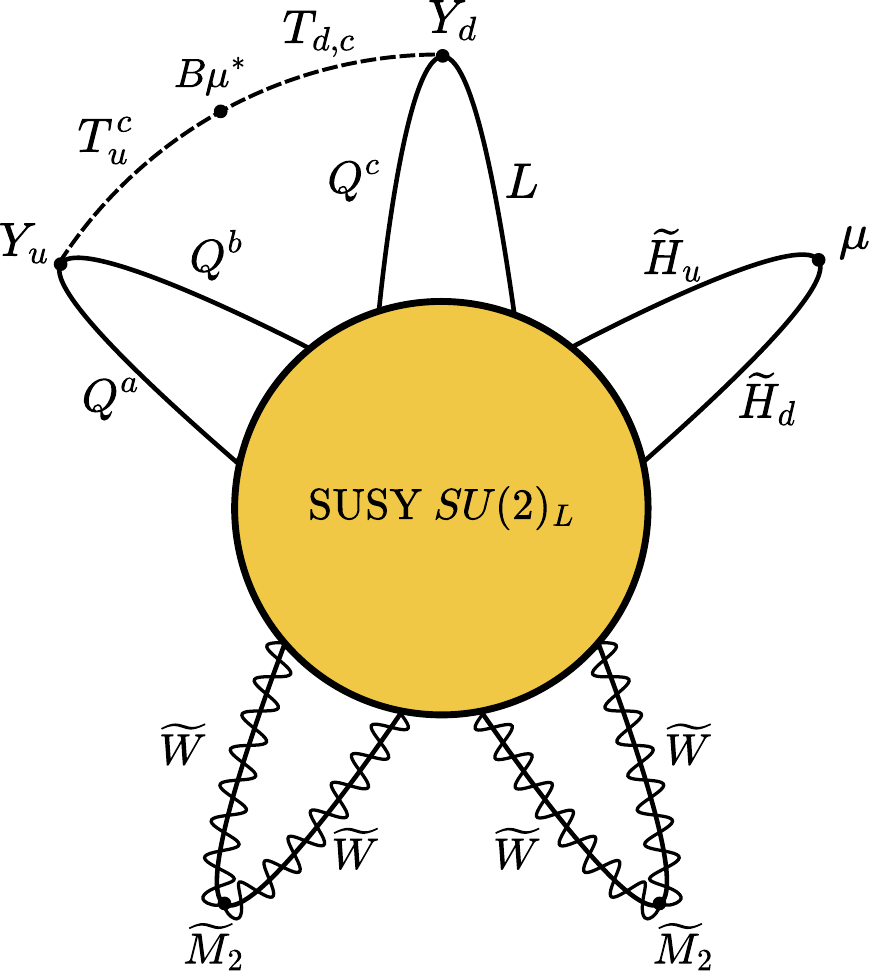}
         \caption{}
         \label{sm-like su2}
     \end{subfigure}
%     \hfill 
\quad \quad \quad \quad
     \begin{subfigure}[b]{0.35\textwidth}
         \includegraphics[width=\textwidth]{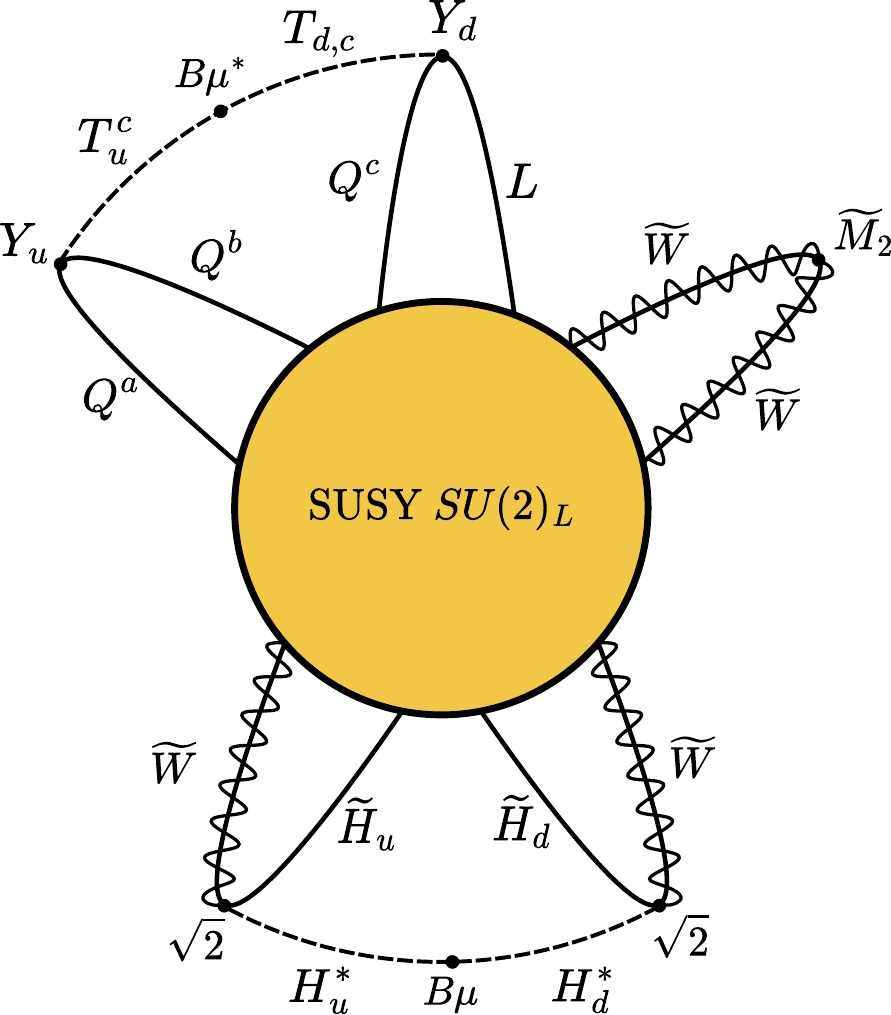}
         \caption{}
         \label{pure susy su2}
     \end{subfigure}
        \caption{Supersymmetric $SU(2)_L$ instanton-induced vacuum diagrams.}
        \label{fig:SUSYSU3inst}
        \end{center}
\end{figure}

The vacuum-to-vacuum amplitude is obtained from Eq. $\eqref{vacuum to vacuum amplitude in interacting theory}$ by expanding the exponential at the lowest order in the couplings, resulting in two contributions. The first one is
\begin{align}
    \mathcal{Z}_{SU(2)_L}^{(a)}=&~e^{-i\theta_{\rm EW}}\int d^4x_0 \int\frac{d\rho}{\rho^5}\updelta_2(\rho)\left(\prod_f\int\frac{d\bar{a}_f}{\sqrt{\bar{\upupsilon}_{0f}}}\right) \rho^2 \left[\frac{\widetilde{M}_2}{2g^2}\int_x \widetilde{W}^{\dagger}\cdot\widetilde{W}^{\dagger}  \right]^2 \rho \left[\mu \int_x \widetilde{H}^{\dagger}_u\cdot \widetilde{H}^{\dagger}_d \right] \rho^{2} B\mu\nonumber\\
    \times & \left[\frac{Y_u Y_d}{2}\int_{\{x_i\}} \varepsilon^{i_1 j_1}Q^{\dagger}_{a_1 i_1}(x_1)L^{\dagger}_{j_1}(x_1)  \left[D_{T_d}(x_1,x_3) \right]^{a_1}_{\,\,\,b_1 }\left[D_{T_u}(x_3,x_2) \right]^{b_1}_{\,\,\, c_2 }   \varepsilon^{a_2b_2c_2}\varepsilon^{i_2j_2}Q^{\dagger}_{a_2i_2}(x_2) Q^{\dagger}_{b_2j_2}(x_2)\right]\, .
\end{align}
where the instanton size $\rho$ is integrated between $M_{\rm GUT}^{-1}$ and $\widetilde{M}_S^{-1}$, $\mathcal{K}_{\alpha}=1$ as we are working in a supersymmetric theory, and, for clarity, we denote the zero modes by their corresponding field. This expression can be illustrated using a 't Hooft diagram, as shown in Figure \ref{sm-like su2}.
%%%%%%%%%%%%%%%%%%%%%%%%%%%%
%%%%%%%%%%%%%%%%%%%%%%%%%%%%
Recalling the expression for the zero modes normalized to unity
\begin{align}
    \left(\widetilde{W}^{\dagger}\right)^{\dot{\alpha}_1 i_1 i_2}=&~f_2(r)\mathcal{A}^{\dot{\alpha}_1\dot{\alpha}_2,i_1 i_2}\left( \frac{2\rho^2}{\pi}\mathcal{M}^{(0)}_{\dot{\alpha}_2}+\frac{\sqrt{2}\rho}{\pi}\left(x\cdot \bar{\sigma}\right)_{\dot{\alpha}_2}^{\,\,\, \beta_2}\mathcal{M}^{(1/2)}_{\beta_2} \right)\, , \\
    \left(\uppsi^{\dagger}\right)^{\dot{\alpha}_1}_{i_1}=&~\frac{\rho}{\pi}f_1(r) \delta^{ \dot{\alpha}_1}_{i_1}\mathcal{M}^{(0)}_{\uppsi^{\dagger}},\qquad \text{for} \qquad \uppsi^{\dagger}=Q^{\dagger}_a,\, L^{\dagger},\, \widetilde{H}_{u,d}^{\dagger}\, ,
\end{align}
we obtain
\begin{align}
    \mathcal{Z}_{SU(2)_L}^{(a)}=&~e^{-i\theta_{\rm EW}}\int d^4x_0 \int\frac{d\rho}{\rho^5}\updelta_2(\rho)(\rho\widetilde{M}_2)^2(\rho\mu)\left(\rho^2 B\mu\right) \left(\prod_{a=1}^3\int d\mathcal{M}_{Q^{\dagger a}}\right)\nonumber\\
    \times &\left[ \frac{Y_u Y_d}{2} \frac{4\rho^4}{\pi^4}\int_{\{x_i \}}f_1^2(x_1)f_1^2(x_2)\mathcal{M}_{Q^{\dagger}_{a_1}}\varepsilon^{a_2b_2c_2}\mathcal{M}_{Q^{\dagger}_{a_2}}\mathcal{M}_{Q^{\dagger}_{b_2}}\left[D_{T_d}(x_1,x_3) \right]^{a_1}_{\,\,\, b_1}\left[D_{T_d}(x_3,x_2) \right]^{b_1}_{\,\,\, c_2}\right]\nonumber\\
    =&~ 3 Y_u Y_d\, e^{-i\theta_{\rm EW}} \int d^4x_0\int \frac{d\rho}{\rho^5}\updelta_2(\rho)(\rho\widetilde{M}_2)^2(\rho\mu)\left(\rho^2 B\mu\right)\left[\frac{4\rho^4}{\pi^4}\int_{\{x_i\}}\frac{D_{T_d}(x_1-x_3)D_{T_u}(x_3-x_2)}{(x_1^2+\rho^2)^3(x_2^2+\rho^2)^3} \right]\, ,
\end{align}
where we used that in the background of an $SU(2)_L$ instanton the propagators of the color triplets are the usual ones
\begin{equation}
    \left[D_{T_{u,d}}(x_1,x_2)\right]^{a}_{\,\,\, b}=\int\frac{d^4p}{(2\pi)^4}\frac{ e^{ip\cdot (x_1-x_2)}}{p^2+m_{T_{u,d}}^2+i\epsilon}\delta^{a}_b\, .
\end{equation}
The generalization to any number of generation of quarks and leptons is straightforward and we have
\begin{align}
    \mathcal{Z}_{SU(2)_L}^{(a)}=&~3^{n_g}(n_g !)\text{det}(Y_u Y_d)_{2n_g}e^{-i\theta_{\rm EW}}\int d^4x_0\int\frac{d\rho}{\rho^5}\updelta_2(\rho)\left(\rho\widetilde{M}_2\right)^2(\rho\mu)\left(\rho^2 B\mu\right)^{n_g} \left[J(m_{T_u},m_{T_d})\right]^{n_g}\, ,
    \label{first SU(2)L diagram}
\end{align}
where we introduced the notation
\begin{equation}
    J(m_{T_u},m_{T_d})=\frac{4\rho^4}{\pi^4}\int d^4 x_1 \int d^4 x_2 \int d^4 x_3\frac{D_{T_d}(x_1-x_3)D_{T_u}(x_3-x_2)}{(x_1^2+\rho^2)^3(x_2^2+\rho^2)^3}\, .
\end{equation}
This expression can be further simplified using the explicit expression of the color triplets propagators
\begin{equation}
    J(m_{T_u},m_{T_d})=\frac{1}{8\pi^2}\int_0^{+\infty}dy\frac{y^5 K_1(y)^2}{(y^2+\rho^2m^2_{T_u})(y^2+\rho^2m^2_{T_d})}\, ,
\end{equation}
where, following \cite{Csaki:2019vte}, we have introduced the modified Bessel functions of the second kind
\begin{equation}
    \int d^4x \frac{e^{-ip\cdot x}}{(x^2+\rho^2)^3}=\frac{\pi^2}{2\rho^2}(p\rho)K_1(p\rho)\, .
\end{equation}
Bounds from proton decay place the color triplets around $10^{17}$~GeV~\cite{Murayama:2001ur,SNO:2022trz,Super-Kamiokande:2020wjk,Super-Kamiokande:2016exg,Super-Kamiokande:2017gev}. Assuming that both triplets share the same mass $m_T$, we can integrate them out at tree-level from this expression to obtain
\begin{align}
    J(m_{T_u},m_{T_d})\simeq &~\frac{1}{8\pi^2}\frac{1}{ m_{T}^4\rho^4}\int_0^{m_T\rho} dy y^5 K_1(y)^2\nonumber\\
    \simeq & ~\frac{1}{5\pi^2 m_T^4 \rho^4}-\frac{e^{-2 m_T \rho}}{32\pi}\left(1+\frac{11}{4 m_T \rho }+\frac{129}{32   m_T^2 \rho ^2}+\frac{531}{128 m_T^3 \rho ^3}+\frac{3843}{2048 m_T^4 \rho ^4} \right)\, .
\end{align}
Thus, Eq. $\eqref{first SU(2)L diagram}$ becomes
\begin{align}
    \mathcal{Z}_{SU(2)_L}^{(a)}=&~3^{n_g}(n_g !)\text{det}(Y_u Y_d)_{2n_g}e^{-i\theta_{\rm EW}}\int d^4x_0\int^{\widetilde{M}_S^{-1}}_{m_T^{-1}}\frac{d\rho}{\rho^5}\updelta_2(\rho)\left(\rho\widetilde{M}_2\right)^2(\rho\mu)\left(\rho^2 B\mu\right)^{n_g}\nonumber\\
    \times &  \left[ \frac{1}{5\pi^2 m_T^4 \rho^4}-\frac{e^{-2 m_T \rho}}{32\pi}\left(1+\frac{11}{4 m_T \rho }+\frac{129}{32   m_T^2 \rho ^2}+\frac{531}{128 m_T^3 \rho ^3}+\frac{3843}{2048 m_T^4 \rho ^4} \right) \right]^{n_g}\, ,
\end{align}
where we have explicitly specified the integration bounds for the instanton size, accounting for the UV cutoff resulting from integrating out the triplets. This expression can be evaluated for given values of $n_g$, using the relation $\updelta_2(\rho)=(\rho M)^{b_{SU(2)_L}^{(0)}}\updelta_2(M^{-1})$ for a given mass scale $M$, where $b_{SU(2)_L}^{(0)}$ denotes the $\beta$-function coefficient of the Supersymmetric $SU(2)_L$ theory. For $n_g=3$, we have parametrically
\begin{equation}
    \mathcal{Z}_{SU(2)_L}^{(a)}\sim \frac{\text{det}(Y_u Y_d)_6}{(8\pi^2)^3} e^{-i\theta_{\rm EW}}V_4 \updelta_2(\widetilde{M}_S^{-1})\mu \widetilde{M}_2^2 \widetilde{M}_S\left(\frac{\widetilde{M}_S}{m_T}\right)^9\left(\frac{B\mu}{\widetilde{M}_S^2} \right)^{n_g}\, .
\end{equation}
The second contribution, illustrated in Figure \ref{pure susy su2}, is as follows
\begin{align}
    \mathcal{Z}^{(b)}_{SU(2)_L}=&~e^{-i\theta_{\rm EW}}\int d^4x_0\int\frac{d\rho}{\rho^5}\updelta_2(\rho)\left( \prod_i \frac{d\bar{a}_i}{\sqrt{\bar{\upupsilon}_{0i}}}\right)\rho\left[\frac{\widetilde{M}_2}{2g^2}\int_x \widetilde{W}^{\dagger}\cdot\widetilde{W}\right]\left|\rho^2 B\mu \right|^2 \frac{Y_u Y_d}{2}\nonumber\\
    \times & \left[\int_{\{x_i\}} \varepsilon^{i_1 j_1}Q^{\dagger}_{a_1 i_1}(x_1)L^{\dagger}_{j_1}(x_1)  \left[D_{T_d}(x_1,x_3) \right]^{a_1}_{\,\,\,b_1 }\left[D_{T_u}(x_3,x_2) \right]^{b_1}_{\,\,\, c_2 }   \varepsilon^{a_2b_2c_2}\varepsilon^{i_2j_2}Q^{\dagger}_{a_2i_2}(x_2) Q^{\dagger}_{b_2j_2}(x_2)\right]\nonumber\\
    \times & \left[2\int_{\{y_i\}}\widetilde{H}_u^{\dagger}(y_1)\cdot\widetilde{W}^{\dagger}(y_1)\cdot  \left[D_{H_u}(y_1,y_3) \right]\cdot\left[ D_{H_d}(y_3,y_2)\right]\cdot \widetilde{W}^{\dagger}(y_2)\cdot\widetilde{H}_d^{\dagger}(y_2) \right]\, .
\end{align}
This can only be computed in the limit where the Higgs fields are massless, as the propagator for massive particles charged under the gauge group associated to the background instanton is not known \cite{Brown:1978bta,Din:1979fh}. Parametrically, we find that the contribution to the axion potential is subdominant compared to $\mathcal{Z}^{(a)}_{SU(2)_L}$. However, to compute this contribution, one would need to use the Green's function for a scalar field charged under $SU(2)_L$ in the background of an $SU(2)_L$ instanton, given in regular gauge by \cite{Brown:1977eb}
\begin{equation}
    D^{i}_{\,\,\, j}(x,y)=\frac{(\rho^2+|x|\,|y|)\delta_j^{i}+2i(\bar{\sigma}_{\mu\nu})^{i}_{\,\,\, j}x_{\mu}y_{\nu}}{(x^2+\rho^2)^{1/2}4\pi^2(x-y)^2 (y^2+\rho^2)^{1/2}}\, ,
    \label{complicated propagator scalar field}
\end{equation}
such a computation, however, is beyond the scope of this paper.

\subsection{SQCD $+$ color triplets $T_{u,d}$}

We consider this example because of its non-trivial aspect due to the presence of the color triplets $T_u$ and $T_d$. For a subset of diagrams we can perform the computation even in the case where the massive color triplets are propagating. We consider the following superpotential
\begin{align}
    W_{SU(3)_c}\supset Y_u\left[\frac{1}{2}\varepsilon_{ij}\varepsilon_{abc}\widetilde{Q}^{ai}\widetilde{Q}^{bj}T_u^c-\varepsilon_{ij}\widetilde{Q}^{ai}\widetilde{\bar{u}}_a H_u^{j} \right]+ Y_d \left[\varepsilon^{abc}\widetilde{\bar{u}}_a\widetilde{\bar{d}}_b T_{d,c}+\widetilde{Q}^{ai}\widetilde{\bar{d}}_a H_{d,i} \right]+\mu_T T_u T_d\, .
\end{align}
In addition to the superpotential, we also account for the Yukawa-gauge interactions for each chiral supermultiplet and include the following soft SUSY-breaking terms
\begin{equation}
    \mathcal{L}_{\rm soft}\supset -\frac{1}{2g^2}\widetilde{M}_3 \tilde{g}\tilde{g}-B\mu\left(T_u T_d+H_u H_d \right)+\text{h.c.}\, .
\end{equation}
\begin{figure}[!t]
\begin{center}
\includegraphics[width=1\textwidth]{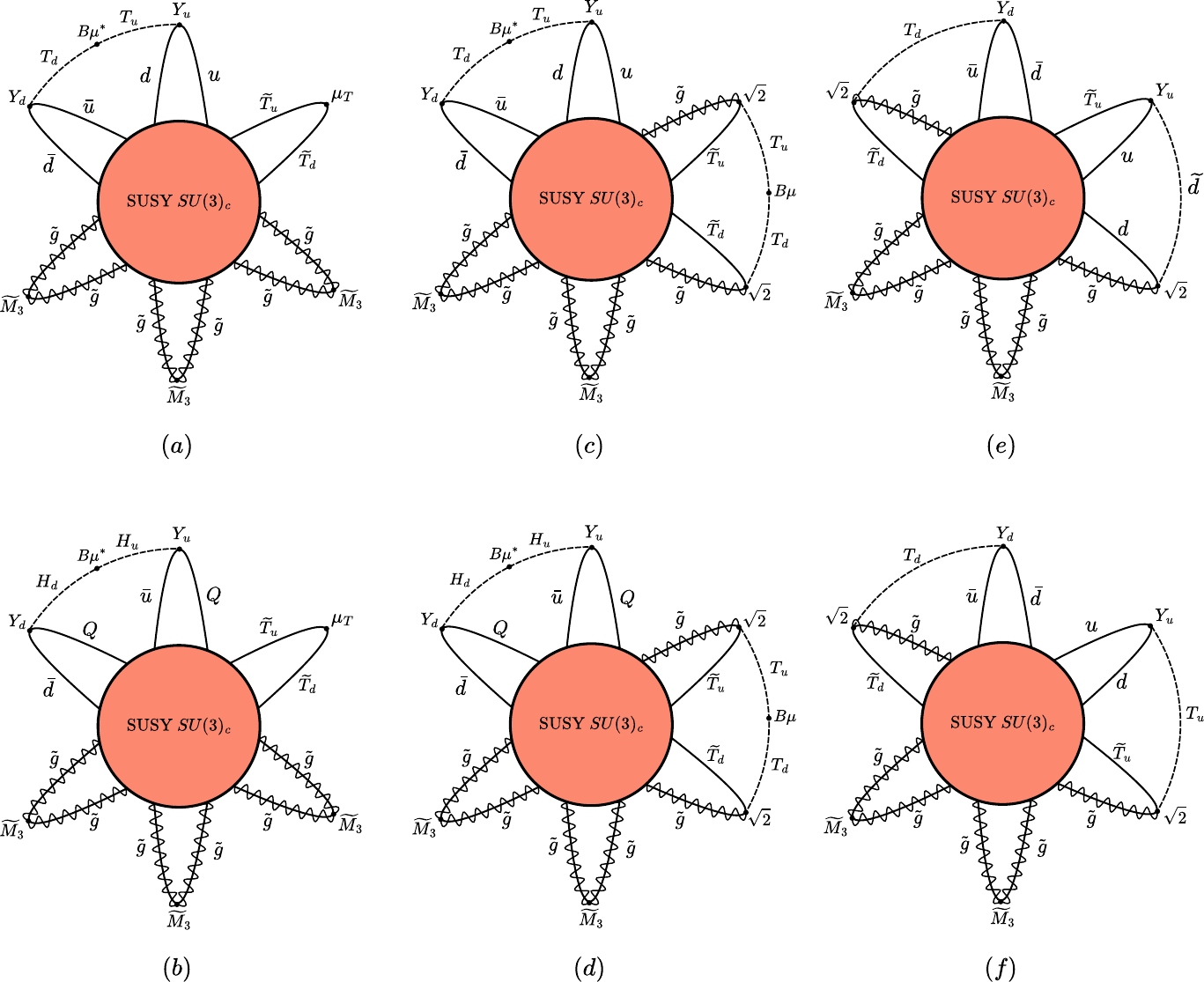}
\caption{Examples of SQCD instanton-induced vacuum diagrams in presence of color triplets $T_{u,d}$.}
\label{fig:SQCDdiagrams}
\end{center}
\end{figure}
For clarity, we will focus on one generation of quarks and leptons as the generalization to any generation is straightforward from these results. In this case, the diagram in Figure \ref{fig:SQCDdiagrams}a gives the following contribution to the vacuum energy
\begin{align}
    \mathcal{Z}^{(a)}_{SU(3)_c}=&~Y_u Y_d e^{-i\theta}\int d^4x_0 \int\frac{d\rho}{\rho^5}\updelta_3(\rho)(\rho \widetilde{M}_3)^3(\rho\mu_T)\left(\rho^2 B\mu\right)\int \frac{d\mathcal{M}_{u^{\dagger}}}{\sqrt{\bar{\upupsilon}_u}}
    \frac{d\mathcal{M}_{\bar{u}^{\dagger}}}{\sqrt{\bar{\upupsilon}_{\bar{u}}}} \frac{d\mathcal{M}_{d^{\dagger}}}{\sqrt{\bar{\upupsilon}_d}} \frac{d\mathcal{M}_{\bar{d}^{\dagger}}}{\sqrt{\bar{\upupsilon}_{\bar{d}}}} \nonumber\\
    \times &\left[\int_{\{x_i\}} \varepsilon_{a_1 b_1 c_1}\bar{u}^{\dagger b_1}(x_1) \bar{d}^{\dagger c_1}(x_1) \left[D_{T_d}(x_1,x_3)\right]^{a_1}_{\,\,\, d}\left[D_{T_u}(x_3,x_2)\right]^{d}_{\,\,\, c_2}\varepsilon^{a_2b_2c_2}u^{\dagger}_{a_2}(x_2)d_{b_2}^{\dagger}(x_2) \right]\nonumber\\
    =&~ Y_u Y_d e^{-i\theta} \int d^4 x_0 \int\frac{d\rho}{\rho^5}\updelta_3(\rho)(\rho\widetilde{M}_3)^3(\rho \mu_T)\left(\rho^2 B\mu\right) \left[\frac{4\rho^4}{\pi^4}\int_{\{x_i\}}\frac{\left[D_{T_d}(x_1,x_3)\right]^{3}_{\,\,\, a}\left[D_{T_d}(x_3,x_2)\right]^{a}_{\,\,\, 3}}{(x_1^2+\rho^2)^3(x_2^2+\rho^2)^3} \right]\nonumber\\
    =&~ Y_u Y_d e^{-i\theta} \int d^4 x_0 \int\frac{d\rho}{\rho^5}\updelta_3(\rho)(\rho\widetilde{M}_3)^3(\rho \mu_T)\left(\rho^2 B\mu\right)J(m_{T_u},m_{T_d})\, .
\end{align}
We see that even if the propagating scalar field is charged under the instanton's gauge group, we can still compute it with the usual massive propagator, as only the singlet part of the propagator with respect to the instanton corner is selected in this calculation. The same analysis conducted for the $SU(2)_L$ case can also be applied to this contribution to the vacuum-to-vacuum amplitude.

Now, the diagram with Higgs doublet loops in Figure \ref{fig:SQCDdiagrams}b gives
\begin{equation}
    \mathcal{Z}^{(b)}_{SU(3)_c}=2 Y_u Y_de^{-i\theta}\int d^4x_0 \int\frac{d\rho}{\rho^5}\updelta_3(\rho)(\rho\widetilde{M}_3)^3 (\rho\mu_T)\left(\rho^2 B\mu\right)J(m_{H_u},m_{H_d})\, ,
    \label{SQCD doublet loop}
\end{equation}
where the factor of $2$ arises from the decomposition of this diagram into two separate diagrams, each corresponding to one component of $Q=(u,d)$. To simplify this expression, we assume that one of the doublets, specifically $H_d$, is at the SUSY-breaking scale $\widetilde{M}_S$, while $H_u$ is at the EW scale. Thus, considering that $m_{H_u}\ll m_{H_d}=\widetilde{M}_S$, we find that
\begin{equation}
    J(m_{H_u},m_{H_d})\simeq \frac{1}{8\pi^2}\log\left[\frac{2e^{-\left(\frac{1}{2}+\gamma_E\right)}}{m_{H_d}\rho}\right]\, .
\end{equation}
With this simplification, Eq. $\eqref{SQCD doublet loop}$ becomes
\begin{align}
     \mathcal{Z}^{(b)}_{SU(3)_c}=2 Y_u Y_de^{-i\theta}\int d^4x_0 \int_{M_{\rm GUT}^{-1}}^{\widetilde{M}_S^{-1}}\frac{d\rho}{\rho^5}\updelta_3(\rho)(\rho\widetilde{M}_3)^3 (\rho\mu_T)\left(\rho^2 B\mu\right)\frac{1}{8\pi^2}\left[\frac{2e^{-\left(\frac{1}{2}+\gamma_E\right)}}{m_{H_d}\rho}\right]\, .
\end{align}

In the expansion of the exponential in Eq. $\eqref{vacuum to vacuum amplitude in interacting theory}$, we find that additional diagrams are generated, categorized into two types. The first type mirrors the diagram shown in Figure \ref{pure susy su2}, and is represented in Figures \ref{fig:SQCDdiagrams}c and \ref{fig:SQCDdiagrams}d. Although these diagrams are not fully computable, they are parametrically subdominant compared to those we have just computed. For reference, we present their formal expression
\begin{align}
    \mathcal{Z}_{SU(3)_c}^{(c)}=&~Y_u Y_d e^{-i\theta}\int d^4x_0\int\frac{d\rho}{\rho^5}\updelta_3(\rho)\left(\prod_i\frac{d\bar{a}_i}{\sqrt{\bar{\upupsilon}_{0i}}}\right)\rho^2\left[\frac{\widetilde{M}_3}{2 g_s}\int_x \tilde{g}^{\dagger}\cdot\tilde{g}^{\dagger} \right]^2|\rho^2 B\mu|^2\nonumber\\
    \times & \left[\int_{\{x_i \}} \varepsilon_{a_1 b_1 c_1}\bar{u}^{\dagger b_1}(x_1) \bar{d}^{\dagger c_1}(x_1) \left[D_{T_d}(x_1,x_3)\right]^{a_1}_{\,\,\, d}\left[D_{T_u}(x_3,x_2)\right]^{d}_{\,\,\, c_2}\varepsilon^{a_2b_2c_2}u^{\dagger}_{a_2}(x_2)d_{b_2}^{\dagger}(x_2) \right]\nonumber\\
    \times & \left[2\int_{\{y_i\}}\widetilde{T}_u^{\dagger}(x_1)\cdot\tilde{g}^{\dagger}(x_1)\cdot\left[D_{T_u}(x_1,x_3) \right]\cdot \left[D_{T_d}(x_3,x_2)\right]\cdot \tilde{g}^{\dagger}(x_2)\cdot\widetilde{T}_d^{\dagger}(x_2) \right]\, ,
\end{align}
and the expression of $\mathcal{Z}_{SU(3)_c}^{(d)}$ is obtained from the previous result by substituting the triplets loops closing the quarks legs with loops involving Higgs doublets.
\begin{comment}
\begin{align}
    \mathcal{Z}_{SU(3)_c}^{(d)}=&Y_u Y_d e^{-i\theta}\int d^4x_0\int\frac{d\rho}{\rho^5}\updelta_3(\rho)\left(\prod_i\frac{d\bar{a}_i}{\sqrt{\bar{\upupsilon}_{0i}}}\right)\rho^2\left[\frac{\widetilde{M}_3}{2 g_s}\int_x \tilde{g}^{\dagger}\cdot\tilde{g}^{\dagger} \right]^2|\rho^2 B\mu|^2\nonumber\\
    \times & \left[\int_{\{x_i \}}\bar{d}^{\dagger a}(x_1) Q^{\dagger}_{ai_1}(x_1)\left[D_{H_d}(x_1,x_3) \right]^{i_1}_{\,\,\, i_2}\left[D_{H_u}(x_3,x_2) \right]^{i_2}_{\,\,\, i_3}\varepsilon^{i_3 i_4}Q^{\dagger}_{b i_4}\bar{u}^{\dagger b} \right]\nonumber\\
    \times & \left[2\int_{\{y_i\}}\widetilde{T}_u^{\dagger}(x_1)\cdot\tilde{g}^{\dagger}(x_1)\cdot\left[D_{T_u}(x_1,x_3) \right]\cdot \left[D_{T_d}(x_3,x_2)\right]\cdot \tilde{g}^{\dagger}(x_2)\cdot\widetilde{T}_d^{\dagger}(x_2) \right]
\end{align}
\end{comment}
Along with these diagrams, the expansion of the exponential in Eq. $\eqref{vacuum to vacuum amplitude in interacting theory}$ yields nine additional diagrams. Two of these are depicted in Figures \ref{fig:SQCDdiagrams}e and \ref{fig:SQCDdiagrams}f, while the remaining seven are obtained through permutations of the fields within the loops. Although these diagrams involve propagating scalar fields charged under the instanton's gauge group, they are fully computable, as we will demonstrate with the following example. The first diagram results in
\begin{align}
    \mathcal{Z}^{(e)}_{SU(3)_c}=&~ 2 Y_u Y_d e^{-i\theta} \int d^4x_0 \int\frac{d\rho}{\rho^5}\updelta_3(\rho)\rho^6\left(\prod_i\frac{d\bar{a}_i}{\sqrt{\bar{\upupsilon}_{0i}}}\right)\left[\frac{\widetilde{M}_3}{2 g^2}\int_x \tilde{g}^{\dagger}\cdot\tilde{g}^{\dagger} \right]^2\nonumber\\
    \times & \left[\int_{x_1,x_2 }d^{\dagger}_{a_1}(x_1)\tilde{g}^{\dagger a_1 }_{\,\,\,\,\, b_1}(x_1)\left[D_{\widetilde{d}}(x_1,x_2) \right]^{b_1}_{\,\,\, b_2}   \varepsilon^{a_2b_2c_2}u^{\dagger}_{a_2}(x_2)\widetilde{T}_{u,c_2}^{\dagger}(x_2)\right]\nonumber\\
    \times & \left[\int_{x_3,x_4 }\varepsilon_{a_3b_3c_3}\bar{u}^{\dagger a_3}(x_3) \bar{d}^{\dagger b_3}(x_3)\left[D_{T_d}(x_3,x_4) \right]^{c_3}_{\,\,\,\,\, c_4}\tilde{g}^{c_4}_{\,\,\, a_4}(x_4)  \widetilde{T}_{d}^{\dagger a_4}(x_4) \right]\nonumber\\
    =& ~Y_u Y_d e^{-i\theta} \int d^4x_0\int\frac{d\rho}{\rho^5}\updelta_3(\rho)(\rho\widetilde{M}_3)^2\rho^4 I(m_{\widetilde{d}})I(m_{T_d})\, ,
    \label{first diagram pure SUSY SQCD}
\end{align}
where we introduced the notation
\begin{equation}
    I(m_\phi)=\frac{4\rho^4}{\pi^4}\int_{x_1,x_2}\frac{D_{\phi}(x_1-x_2)}{(x_1^2+\rho^2)^3(x_2^2+\rho^2)^3}\, .
\end{equation}
 Once again, we observe that the propagator of scalar fields charged under the instanton's gauge group reduces to the standard form. The remaining diagrams in this family are derived from Eq. $\eqref{first diagram pure SUSY SQCD}$ by substituting the propagators with the corresponding ones. All these diagrams share the same parametric dependence and overall sign, differing only in the specific particles propagating within the loops. For instance, the diagram shown in Figure \ref{fig:SQCDdiagrams}f is given by
\begin{equation}
    \mathcal{Z}_{SU(3)_c}^{(f)}=Y_u Y_d e^{-i\theta} \int d^4x_0 \int\frac{d\rho}{\rho^5}\updelta_3(\rho)(\rho\widetilde{M}_3)^2 \rho^4 I(m_{T_u}) I(m_{T_d})\, .
\end{equation}
This completes the analysis of instanton contributions to the vacuum energy in SQCD extended by the inclusion of two color triplets, $T_{u,d}$.

\subsection{Minimal SUSY $SU(5)$}

In this section we consider the minimal Supersymmetric $SU(5)$ theory without its symmetry breaking sector. The theory has the following superpotential
\begin{equation}
W_{SU(5)}\supset Y_5 \Phi_{\bar{5},i}\Phi_{10}^{ ij}H_{\bar{5},j}+\frac{1}{8}Y_{10}\varepsilon_{ijklm}\Phi_{10}^{ij}\Phi_{10}^{kl}H_{5}^m+\mu_5 H_{\bar{5},i}H_5^{i}\, .
\end{equation}
In addition to the superpotential we consider the Yukawa-gauge interactions, and the SUSY-breaking terms
\begin{equation}
\mathcal{L}_{\rm soft}\supset -\frac{1}{2g_5^2}\widetilde{M}_5 \lambda_5 \lambda_5 -B\mu H_{\bar{5}}H_{5}+\text{h.c.}\, .
\end{equation}
In the background of an instanton the fermion zero mode content of the theory is the following: $\lambda^{\dagger}_5$ has $2T(\textbf{Adj})=10$ zero modes, $\Psi_{10}^{\dagger}$ has $2T(\textbf{10})=3$, $\Psi_{\bar{5}}^{\dagger}$, $\widetilde{H}_{\bar{5}}^{\dagger}$ and $\widetilde{H}_{5}^{\dagger}$ have $2T(\textbf{Fund})=1$ each. In Eq. $\eqref{fermion zero mode antisymmetric}$ we obtained the expression of the fermion zero mode in the antisymmetric representation for $SU(N)$, specializing to $N=5$ and normalizing it to one, we obtain
\begin{equation}
    \Psi_{10,ij}^{\dagger \dot{\alpha}}=\frac{\rho}{\pi}f_1(r)\left[\delta^{\dot{\alpha}}_i\upmu_{j}-\delta^{\dot{\alpha}}_j\upmu_{i} \right],\qquad 
    \upmu=
    \begin{pmatrix}
        0 && 0 && \upmu_{3} && \upmu_{4} && \upmu_{5}\\
    \end{pmatrix}\, .
\end{equation}
The vacuum-to-vacuum amplitude is derived from Eq. $\eqref{vacuum to vacuum amplitude in interacting theory}$ by expanding the exponential to the lowest order in the couplings needed to saturate the Grassmann integrations. The first contribution is
\begin{align}
    \mathcal{Z}_{SU(5)}^{(a)}= &~ e^{-i\theta}\int d^4x_0\int\frac{d\rho}{\rho^5}\updelta_5(\rho)\left(\prod_i\int\frac{d\bar{a}_i}{\sqrt{\bar{\upupsilon}_{0i}}}\right)\rho^{5}\left[\frac{\widetilde{M}_5}{2g_5^2}\int_x \lambda_5^{\dagger}\cdot\lambda_5^{\dagger} \right]^5\rho\left[\mu_5\int_x \widetilde{H}_{5}^{\dagger}\cdot \widetilde{H}^{\dagger}_{\bar{5}} \right]\rho^2 B\mu \frac{Y_5 Y_{10}}{8}\nonumber\\
    \times & \left[\int_{\{ x_i \}} \Psi^{\dagger i_1}_{\bar{5}}(x_1)\Psi_{10,i_1j_1}^{\dagger}(x_1) \left[D_{H_{\bar{5}}}(x_1,x_3) \right]^{j_1}_{\,\,\, k}\left[D_{H_{5}}(x_3,x_2) \right]^{k}_{\,\,\, m_2}\varepsilon^{i_2j_2k_2l_2m_2}\Psi_{10,i_2j_2}^{\dagger}(x_2)\Psi_{10,k_2l_2}^{\dagger}(x_2) \right]\, .
\end{align}
Using the expression of the fermion zero modes, this simplifies to
\begin{align}
    \mathcal{Z}_{SU(5)}^{(a)}=& ~Y_5 Y_{10} e^{-i\theta}\int d^4 x_0 \int\frac{d\rho}{\rho^5}\updelta_5(\rho)(\rho\widetilde{M}_5)^5(\rho\mu_5)(\rho^2 B\mu)\left(\prod_{u=3}^{5}\int\frac{d\upmu_u}{\sqrt{\bar{\upupsilon}_{10}}}\right)\nonumber\\
    \times & \left[\frac{4\rho^4}{\pi^4}\int_{\{x_i\}}f^2_1(x_1)f_1^2(x_2)\upmu_3\upmu_4\upmu_5\sum_{u=3}^5\Big(\left[D_{H_{\bar{5}}}(x_1,x_3) \right]^{u}_{\,\,\, v}\left[D_{H_{5}}(x_3,x_2) \right]^{v}_{\,\,\, u} \Big)\right]\, ,
\end{align}
where the sum over $u$ selects only the singlet part of the propagators with respect to the instanton corner, we thus have
\begin{align}
\mathcal{Z}^{(a)}_{SU(5)}=3Y_5 Y_{10}\int d^4x_0 \int\frac{d\rho}{\rho^5}\updelta_5(\rho)(\rho \widetilde{M}_5)^5 (\rho \mu_5) (\rho^{2}B\mu)J(\rho,m_5,m_{\bar{5}})\, .
\end{align}
As in the case of SQCD with triplets $T_{u,d}$, there are additional diagrams we have to take into account. They are of the same kind of those previously exposed. 
%%%%%%%%%%%%%%%%%%%%%%%%%%%%
\begin{figure}[!t]
\begin{center}
\includegraphics[width=0.8\textwidth]{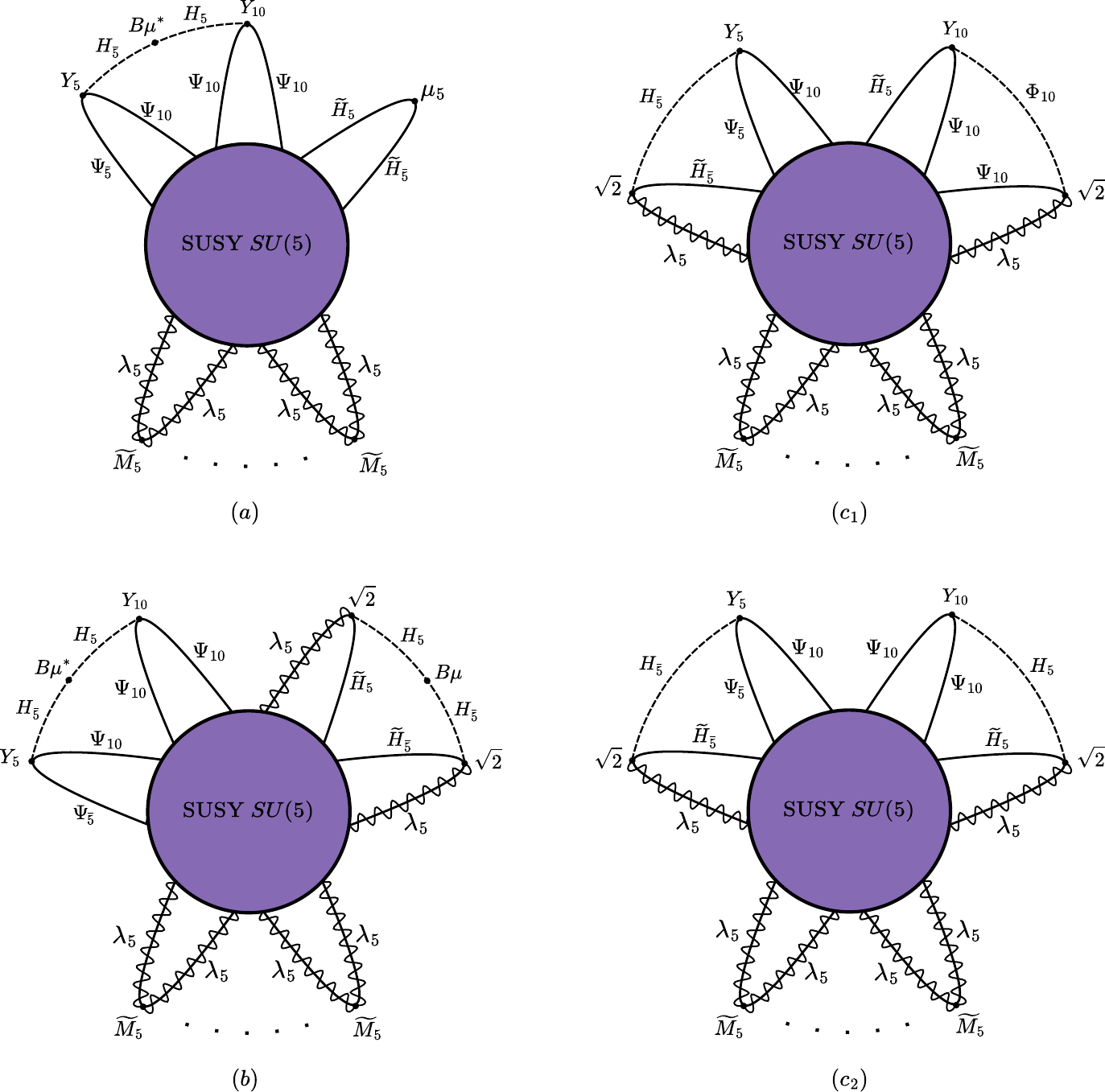}
\caption{Examples of Supersymmetric $SU(5)$ instanton-induced vacuum diagrams. There are $10$ gaugino legs in the diagram from their $SU(5)$ Dynkin index.}
\label{fig:SUSYSU5inst}
\end{center}
\end{figure}
%%%%%%%%%%%%%%%%%%%%%%%%%%%%
Again, the expansion of the exponential in Eq. $\eqref{vacuum to vacuum amplitude in interacting theory}$ yield a diagram that is not fully computable due to the propagation of massive scalar fields charged under the instanton's gauge group. For reference, we give its formal expression
\begin{align}
    \mathcal{Z}_{SU(5)}^{(b)}= & ~e^{-i\theta}\int d^4x_0\int\frac{d\rho}{\rho^5}\updelta_5(\rho)\left(\prod_i\frac{d\bar{a}_i}{\sqrt{\bar{\upupsilon}_{0i}}}\right)\rho^4\left[\frac{\widetilde{M}_5}{2g_5}\int_x\lambda^{\dagger}_5\cdot\lambda_5^{\dagger} \right]^4|\rho^2 B\mu|^2 \frac{Y_5 Y_{10}}{8}\nonumber\\
    \times & \left[\int_{\{ x_i \}} \Psi^{\dagger i_1}_{\bar{5}}(x_1)\Psi_{10,i_1j_1}^{\dagger}(x_1) \left[D_{H_{\bar{5}}}(x_1,x_3) \right]^{j_1}_{\,\,\, k}\left[D_{H_{5}}(x_3,x_2) \right]^{k}_{\,\,\, m_2}\varepsilon^{i_2j_2k_2l_2m_2}\Psi_{10,i_2j_2}^{\dagger}(x_2)\Psi_{10,k_2l_2}^{\dagger}(x_2) \right]\nonumber\\
    \times & \left[2\int_{\{y_i\}} \widetilde{H}_5^{\dagger}(x_1)\cdot\lambda_5^{\dagger}(x_1)\cdot\left[D_{H_5}(x_1,x_3) \right]\cdot \left[D_{H_{\bar{5}}}(x_3,x_2)\right]\cdot\lambda^{\dagger}_5(x_2)\cdot\widetilde{H}_{\bar{5}}^{\dagger}(x_2) \right]\, .
\end{align}
This expression requires the expression of the propagator of massive scalar fields charged under the instanton's gauge group, as the gaugino precisely selects this part of the scalars propagators in the right loop of Figure \ref{fig:SUSYSU5inst}b. This result is only known for the massless case, as given in Eq. $\eqref{complicated propagator scalar field}$. Fortunately, this diagram is parametrically smaller than the contribution from $\mathcal{Z}_{SU(5)}^{(a)}$.
\newline

There are six additional diagrams coming from the expansion of the exponential in Eq. $\eqref{vacuum to vacuum amplitude in interacting theory}$, two of them are shown in Figures \ref{fig:SUSYSU5inst}c$_1$ and  \ref{fig:SUSYSU5inst}c$_2$. The one in Figure \ref{fig:SUSYSU5inst}c$_2$ is given by
\begin{align}
    \mathcal{Z}_{SU(5)}^{(c_2)}=&~e^{-i\theta}\int d^4x_0\int\frac{d\rho}{\rho^5}\updelta_5(\rho)\rho^8\left(\prod_i\frac{d\bar{a}_i}{\sqrt{\bar{\upupsilon}_{0i}}}\right)\left[\frac{\widetilde{M}_5}{2g_5^2}\int_x\lambda^{\dagger}_5\cdot\lambda^{\dagger}_5 \right]^4\nonumber\\
    \times & \left[\sqrt{2}Y_5\int_{x_1,x_2}\Psi_{\bar{5}}^{\dagger i}(x_1) \Psi_{10,ij}^{\dagger}(x_1)\left[D_{H_{\bar{5}}}(x_1,x_2)\right]^{j}_{\,\,\, k} \left(\lambda_5^{\dagger}\right)^{k}_{\,\,\, l}(x_2) \widetilde{H}_{\bar{5}}^{\dagger l}(x_2) \right]\nonumber\\
    \times & \left[\sqrt{2}\frac{Y_{10}}{8}\int_{y_1,y_2} \widetilde{H}_{5,a}^{\dagger}(y_1)\left(\lambda_5^{\dagger} \right)^{a}_{\,\,\, b}(y_2)\left[D_{H_5}(y_1,y_2) \right]^{b}_{\,\,\, m} \varepsilon^{ijklm}\Psi_{10,ij}^{\dagger}(y_2)\Psi_{10,kl}^{\dagger}(y_2)  \right]\, .
\end{align}
Using the expressions of the fermion zero modes, this simplifies to
\begin{align}
    \mathcal{Z}_{SU(5)}^{(c_2)}=Y_5 Y_{10}e^{-i\theta}\int d^4x_0 \int\frac{d\rho}{\rho^5}\updelta_5(\rho)(\rho\widetilde{M}_5)^4 \rho^4 I(m_{H_5})I(m_{H_{\bar{5}}})\, .
\end{align}
The diagram in Figure \ref{fig:SUSYSU5inst}c$_1$ contributes to the vacuum-to-vacuum amplitude as
\begin{align}
    \mathcal{Z}_{SU(5)}^{(c_1)}=2 Y_5 Y_{10}e^{-i\theta}\int d^4x_0 \int\frac{d\rho}{\rho^5}\updelta_5(\rho)(\rho\widetilde{M}_5)^4 \rho^4 I(m_{10})I(m_{H_{\bar{5}}})\, ,
\end{align}
where the factor of $2$ arises from the propagator of $\Phi_{10}$. This highlights that the functional method introduced in this paper enables computations that were previously unattainable with earlier approaches aimed at precisely calculating instanton-induced axion potentials.

%%%%%%%%%%%%%%%%%%%%%%%%%%%%%%%%%%%%%%%%%%%%%%%%
\section{Conclusion}\label{sec:conclusion}

Driven by the growing interest in axion physics and the ubiquitous role of instantons in axion model building, we developped a method to perform precise calculations of instanton-induced axion potential. We have reviewed that, within the the dilute instanton gas approximation, this calculation reduces to the evaluation of the vacuum-to-vacuum amplitude in the background of a single instanton, which was the central object of our analysis.

The method presented in this paper is based on the conventional techniques used to evaluate scattering amplitudes in Quantum Field Theory (QFT) through functional methods. Therefore, we have transformed the often complex task of evaluating contributions to the axion potential into a more manageable process, grounded in standard QFT techniques, thus providing a solid foundation for practical applications.

One of the main subtleties in these computations lies in the treatment of fermions, which cause the amplitude to vanish in the free theory. Obtaining a non-zero result requires the inclusion of interactions. We have constructed the fermionic part of the generating functional, incorporating sources to establish a clear framework to treat interactions. In this sector, the zero modes require a distinct treatment from the non-zero modes. While the non-zero modes are integrated out, we explicitly retain only the zero modes, whose coupling to sources is central to  analysis. Many challenges in these computations arise from the zero modes themselves; for instance, in models with exotic fermion representations, explicit calculations were previously not feasible due to the lack of expressions for these zero modes.
In this work, we constructed the fermion zero modes for arbitrary representations of $SU(2)$, and outlined a procedure to extend these results to any representation of $SU(N)$ within the minimal embedding framework. As a result, we provided a procedure to evaluate the instanton-induced axion potentials in $SU(N)$ Yang-Mills theory with matter content in any representation of the gauge group. To illustrate our methodology, we provided several examples, including the MSSM and the minimal SUSY $SU(5)$ GUT, with sufficient detail to capture all the subtleties involved in these computations.

In conclusion, we have introduced a robust method for computing instanton-induced axion potentials that can be readily applied to a wide range of theories, providing a solid foundation for such analyses.

%%%%%%%%%%%%%%%%%%%%%%%%%%%%%%%%%%%
\section*{Acknowledgments}

I am very grateful to Raffaele Tito D'Agnolo and Marie Sellier-Prono for very useful discussions and comments on the manuscript. I also thank Pier Giuseppe Catinari, Csaba Cs\'aki, Giacomo Ferrante, Eric Kuflik, Stéphane Lavignac, Florian Nortier, Gabriele Rigo and Marcello Romano  for very useful discussions. 
%%%%%%%%%%%%%%%%%%%%%%%%%%%%%%%%%%%%%%%%%%%%%%%%%%%%%%%

\appendix

\section{Conventions and useful formulas}

We choose the Levi-Civita symbols such that $ \epsilon^{1 2} = - \epsilon_{1 2} = 1$, which means $ \epsilon^{\alpha \beta} \epsilon_{\beta \gamma} = \delta^{\alpha}_{\gamma}$. The $SU(2)$ indices are lowered and raised as follows
\begin{equation}
\lambda_{\alpha}=\epsilon_{\alpha \beta} \lambda^{\beta}=\epsilon_{\alpha \beta} \epsilon^{\beta \gamma} \lambda_{\gamma}\, .
\end{equation}
A crucial relation in the spinor formalism is
\begin{equation}
    (\bar{\sigma}^{\mu})^{\dot{\alpha} \alpha} = \epsilon^{\alpha \beta} \epsilon^{\dot{\alpha}\dot{\beta}} (\sigma^{\mu})_{\beta \dot{\beta}}\ ,
\end{equation}
which means that he Euclidean 4-vector of Pauli matrices are given by
\begin{equation}
\left(\sigma_{\mu}\right)_{\alpha\dot{\alpha}}=(\vec{\sigma},i\mathds{1})_{\alpha\dot{\alpha}},\qquad \left(\bar{\sigma}_{\mu}\right)^{\dot{\alpha}\alpha}=(\vec{\sigma},-i \mathds{1})^{\dot{\alpha}\alpha}\, .
\end{equation}
We take the following conventions for our $\sigma_{\mu\nu}$ and $\bar{\sigma}_{\mu\nu}$ matrices\footnote{They differ by a sign from \cite{Martin:1997ns,Dreiner:2008tw}.}
\begin{equation}
\bar{\sigma}_{\mu\nu}=\eta_{a\mu\nu}T^{a}=\frac{1}{4i}\left(\bar{\sigma}_{\mu}\sigma_{\nu}-\bar{\sigma}_{\nu}\sigma_{\mu}\right),\quad \sigma_{\mu\nu}=\bar{\eta}_{a\mu\nu}T^{a}=\frac{1}{4i}\left(\sigma_{\mu}\bar{\sigma}_{\nu}-\sigma_{\nu}\bar{\sigma}_{\mu} \right)\, ,
\end{equation}
where $\eta$ and $\bar{\eta}$ are the so-called 't Hooft symbols introduced in Eq. $\eqref{generators L1 and L2}$ to represent the $SU(2)$ generetors of $O(4)\simeq SU(2)\times SU(2)$; for further details, see Appendix \ref{Angular momenta of the problem}. The $4$-vector Pauli matrices satisfy
\begin{equation}
\left(\bar{\sigma}_{\mu}\right)^{\dot{\alpha}\beta}\left(\sigma_{\nu}\right)_{\beta\dot{\beta}}=\delta^{\dot{\alpha}}_{\dot{\beta}}+2i\eta_{a\mu\nu}\left(S^{a}\right)^{\dot{\alpha}}_{\,\,\, \dot{\beta}}\, , \qquad \left(\sigma_{\mu}\right)^{\alpha\dot{\alpha}}\left(\bar{\sigma}_{\nu}\right)_{\dot{\alpha}\beta}=\delta^{\alpha}_{\beta}+2i\bar{\eta}_{a\mu\nu}\left(S^{a}\right)^{\alpha}_{\,\,\, \beta}\, .
\label{sigmabarsigma}
\end{equation}
We denote the $SU(2)$ generators acting on spinor indices as $\left(S^{a}\right)^{\dot{\alpha}}_{\,\,\, \dot{\beta}}=\left(\frac{\sigma^{a}}{2}\right)^{\dot{\alpha}}_{\,\,\, \dot{\beta}}$ to differentiate them from the other $SU(2)$ generators.
\newline

\noindent A useful identity to note is:
\begin{equation}
\left(\bar{\sigma}_{\mu\nu}\right)^{\dot{\alpha}}_{\,\,\,\dot{\beta}}\left(\bar{\sigma}_{\mu\nu}\right)^{i}_{\,\,\,j}=\varepsilon^{\dot{\alpha}i}\varepsilon_{\dot{\beta}j}+\delta^{\dot{\alpha}}_j\delta^{i}_{\dot{\beta}}\, .
\label{sigmabarmunusigmabarmunu}
\end{equation}

\section{Angular momenta of the problem}\label{Angular momenta of the problem}

In Euclidean space, the theory has a spatial $O(4)$ symmetry which is generated by the following angular momentum written in the coordinate representation
\begin{equation}
\mathcal{J}_{\mu\nu}=-i(x_{\mu}\partial_{\nu}-x_{\nu}\partial_{\mu})=
\begin{pmatrix}
0 && K_3 && -K_2 && -M_1\\
-K_3 && 0 && K_1 && -M_2\\
K_2 && -K_1 && 0 && -M_3\\
M_1 && M_2 && M_3 && 0\\
\end{pmatrix}\, .
\end{equation}
From the commutation relation involving $\mathcal{J}_{\mu\nu}$, we see that its components satisfy the $O(4)$ algebra
\begin{equation}
[K_i,K_j]=i\epsilon_{ijk}K_k,\qquad [M_i,M_j]=i\epsilon_{ijk}M_k,\qquad [K_i,M_j]=i\epsilon_{ijk}M_k\, .
\end{equation}
Locally, we can observe that $O(4)\simeq SU(2)_1\times SU(2)_2$ introducing the two angular momentum operators $L_1^{a}$ and $L_2^{a}$ as follows
\begin{equation}
L_1^{i}=\frac{1}{2}\left(K_i+M_i \right),\qquad L_2^{i}=\frac{1}{2}\left(K_{i}-M_{i} \right)\, .
\end{equation} 
They satisfy the two separate $SU(2)$ algebras
\begin{equation}
[L_1^{i},L_1^{j}]=i\epsilon^{ijk}L_1^{k},\qquad [L_2^{i},L_2^{j}]=i\epsilon^{ijk}L_2^{k},\qquad [L_1^{i},L_2^{j}]=0\, .
\end{equation}
They can be nicely expressed using 't Hooft symbols
\begin{equation}
L_1^{a}=-\frac{i}{2}\eta_{a\mu\nu}x_{\mu}\partial_{\nu},\qquad L_2^{a}=-\frac{i}{2}\bar{\eta}_{a\mu\nu}x_{\mu}\partial_{\nu}\, ,
\label{generators L1 and L2}
\end{equation}
where
\begin{equation}
\eta_{a\mu\nu}=
\begin{cases}
\epsilon_{a\mu\nu},\quad &\mu,\nu=1,2,3\\
-\delta_{a\nu},\quad &\mu=4\\
\delta_{a\mu},\quad &\nu=4\\
0,\quad &\mu=\nu=4\\
\end{cases},\quad \text{and}\quad \bar{\eta}_{a\mu\nu}=
\begin{cases}
\epsilon_{a\mu\nu},\quad &\mu,\nu=1,2,3\\
\delta_{a\nu},\quad &\mu=4\\
-\delta_{a\mu},\quad &\nu=4\\
0,\quad &\mu=\nu=4\\
\end{cases}
\end{equation}
The operator $\mathcal{J}^2$ appears in the Laplace-Beltrami operator present in the $4d$ d'Alembert operator. In $4d$ they are related by
\begin{equation}
\Delta_{S^3}=K_1^2+K_2^2+K_3^2+M_1^2+M_2^2+M_3^2\equiv \frac{1}{2}\mathcal{J}^2=-4L_{1}^2=-4L_{2}^2\, .
\end{equation}
In $d$ dimensions, the Laplace-Beltrami operator is diagonalized by special functions known as higher-dimensional spherical harmonics, as defined in \cite{10.1063/1.527513}. In the Appendix of \cite{Arkani-Hamed:2017jhn}, they express $3d$ spherical harmonics in cartesian form using Pauli matrices and $SU(2)$ tensor representations. Here, we extend this approach to four dimensions, generating $4d$ spherical harmonics in cartesian form through a similar construction. Specifically, we introduce
\begin{equation}
    (x\cdot\bar{\sigma})^{i}_{\,\,\, j}=
    \begin{pmatrix}
        x_3-ix_4 & x_1-i x_2\\
        x_1+i x_2 & -x_3-i x_4\\
    \end{pmatrix}\, ,\quad\text{and}\quad  (x\cdot\bar{\sigma})_{ij}=
    \begin{pmatrix}
        -x_1-ix_2 & x_3+i x_4\\
        x_3-i x_4 & x_1-i x_2\\
    \end{pmatrix}\, .
\end{equation}
Since the Laplace-Beltrami operator is expressed in terms of the squares of $L_1^{a}$ and $L_2^{a}$, diagonalizing it requires studying the irreps of the two $SU(2)$ subgroups of $SO(4)$ generated by these operators. To describe a solution with orbital angular momentum $\ell$, we construct symmetric rank $2\ell$ tensors from tensor products of $\ell$ $(x\cdot \bar{\sigma})$. However, as pointed out in \cite{Arkani-Hamed:2017jhn}, there is no need to explicitly build these tensors, as the relevant information about the eigenfunctions is fully encoded in the following object
\begin{equation}
\varphi_{\ell}=\left[\xi\xi(x\cdot\bar{\sigma})\right]^{\ell}=\xi^{i_1}\xi^{j_1}\cdots \xi^{i_{\ell}}\xi^{j_{\ell}}(x\cdot\bar{\sigma})_{i_1 j_1}\cdots (x\cdot\bar{\sigma})_{i_{\ell} j_{\ell}}\, ,
\end{equation}
where $\xi$ is any $2$-component spinor. This object satisfies the eigenvalue equation of the $4d$ Laplace-Beltrami operator
\begin{equation}
    \Delta_{S^3}\varphi_{\ell}=-\ell(\ell+2)\varphi_{\ell}\, .
\end{equation}
In other words, we have solved the eigenvalue equation for the $4d$ Laplace-Beltrami operator by contracting multiple $(x\cdot \bar{\sigma})$ with a fully symmetric tensor.

\section{One-loop determinants} \label{Instanton density}

In this Appendix we provide a detailed derivation of the non-zero modes part of the $SU(N)$ instanton density in the minimal embedding framework in the presence of scalars and fermions in any representation of the gauge group based on \cite{tHooft:1976snw}.
\newline

In \cite{Csaki:2019vte} such a formula for fermions and scalars in the fundamental representation of $SU(N)$ has been derived, however this is not enough to compute instantons contributions to the axion potential in \textit{e.g.} supersymmetric theories or Grand Unified Theories, where we have gauginos or fermions in the $10$-dimensional representation of $SU(5)$.

The primed determinants contain UV divergences due to the infinite number of eigenvalues that can be arbitrarily large. We make the result converge following 't Hooft with two procedures. We first normalize the functional integral for the one instanton background with the same integral in the absence of background. Then, we regulate the UV divergences using Pauli-Villars regularization scheme, with regularization parameter $\mu$. As a result, we compute the following object
\begin{equation}
\textbf{det'} \mathcal{M}\equiv \frac{\text{det'} \mathcal{M} }{\text{det}(\mathcal{M}+\mu^2)}\frac{\det(\mathcal{M}^0+\mu^2)}{\det \mathcal{M}^0}\, ,
\end{equation}
for gauge fields, ghosts, fermions and scalars. 

\subsection{Minimal embedding into $SU(N)$}\label{embedding}

After background field expanding around the instanton solution we have to deal with the following operators written in the background field gauge
\begin{align}
\left(\mathcal{M}_A\right)_{\mu\nu}=&-2\left(\delta_{\mu\nu}D^2-2i F_{\mu\nu}\right) \, ,\quad \left(\mathcal{M}_{\rm ghost}\right)^{ab}=-(D^2)^{ab}\, , \nonumber \\
\left(\mathcal{M}^{(-)}_{\psi}\right)^{\dot{\alpha}\alpha}=& i (\bar{\sigma}_{\mu})^{\dot{\alpha}\alpha}D_{\mu} \, ,\quad \left(\mathcal{M}^{(+)}_{\psi}\right)^{\alpha\dot{\alpha}}= i (\sigma_{\mu})^{\alpha\dot{\alpha}}D_{\mu} \, ,\quad 
\mathcal{M}_{\phi}= - D^2 \, .
\label{operators}
\end{align}
By performing Gaussian integration over quantum fluctuations, we obtain the product of determinants of these operators, from which the contributions of gauge and fermion zero modes have been extracted
\begin{align}
\left(\textbf{det'} \mathcal{M}_A\right)^{-1/2}\left(\textbf{det'} \mathcal{M}_{\psi}\right)\left(\textbf{det} \mathcal{M}_{\rm ghost}\right)
\left(\textbf{det} \mathcal{M}_{\phi}\right)^{-1}\, .
\end{align}
In the following we derive the expression of the normalized and regulated determinants over non-zero modes. This computation was initially performed by 't Hooft for $SU(2)$ \cite{tHooft:1976rip,tHooft:1976snw}. He found that in the covariant background field gauge, the determinants for the gauge fields, ghosts, scalars and fermions could be expressed by a single formula, with the powers of this formula corresponding to the number of degrees of freedom for each field (for an introduction to background field methods see \cite{Peskin:1995ev}). By leveraging the conformal invariance of the classical theory, 't Hooft computed the determinant for the massless complex scalar field in the isospin-$t$ representation of $SU(2)$
\begin{equation}
\textbf{det} \mathcal{M}_{\phi}=\exp\left[\frac{1}{3}T(t)\ln(\mu\rho)+\alpha(t)\right]\, ,
\label{tHooft formula}
\end{equation}
where $T(t)$ is the Dynkin index of the isospin-$t$ representation
\begin{equation}
T(t)=\frac{1}{3}t(t+1)(2t+1)\, ,
\end{equation}
and $\alpha(t)$ is given by
\begin{equation}
\alpha(t)=2T(t)\left(2\mathcal{R}-\frac{1}{6}\ln 2-\frac{1}{9}-\frac{1}{6}t(t+1)+\frac{1}{2}\sum_{s=1}^{2t+1}\left[s(2t+1-s)\left( s-t-\frac{1}{2}\right)\ln s \right]\right)\, ,
\end{equation}
with
\begin{equation}
\mathcal{R}=\frac{\ln 2\pi+\gamma_E}{12}+\frac{1}{2\pi^2}\sum_{s=1}^{\infty}\frac{\ln s}{s^2}\simeq 0.249\, .
\end{equation}
To apply 't Hooft's result to any $SU(N)$ representation, we follow \cite{Bashilov:1978pav} and embed the instanton solution $A_{\mu}^{SU(2)}$ into an $SU(2)$ subgroup of $SU(N)$, allowing us to express it as follows
\begin{equation}
A_{\mu}^{SU(N)}=\left(A_{\mu}^{SU(2)}\right)^{a}T^{a},\qquad a=1,2,3,\qquad [T^{a},T^{b}]=i\varepsilon_{abc}T^c\, .
\end{equation}
This implies that the $T^{a}$'s must form a closed Lie algebra of $SU(2)$, generating a reducible representation of $SU(2)$ on the basis of irreducible adjoint representations of $SU(N)$. Since every reducible representation of $SU(2)$ is completly reducible we can find a unitary transformation that maps the $T^{a}$'s in a block diagonal form where each block has a definite $SU(2)$ isospin. Thus, we can decompose these $T^{a}$'s as a direct sum of generators of isospin representations $t_i$ of $SU(2)$ that we denote $\tau^{a}(t_i)$, namely
\begin{equation}
T^{a}(\textbf{Fund})=\bigoplus_{i\rightarrow \textbf{Fund}} \tau^{a}(t_i),\qquad \sum_{i\rightarrow \textbf{Fund}} (2t_i+1)=N\, ,
\end{equation}
where $i$ runs over the isospin representations of $SU(2)$ involved in the decomposition of the $T^{a}$'s. This can be generalized for arbitrary representation $\mathbf{R}$ of $SU(N)$ and we have
\begin{equation}
T^{a}(\mathbf{R})=\bigoplus_{i\rightarrow \mathbf{R}} \tau^{a}(t_i),\qquad \sum_{i\rightarrow\textbf{R}} (2t_i+1)=\dim(\mathbf{R})\, .
\label{decomposition rep R}
\end{equation}
This means that the operators written in Eq. $\eqref{operators}$ will break into a block diagonal form corresponding to the isospin representations involved in the decomposition of  the original $SU(N)$ representation of the field. For each of these independent blocks we will be able to use Eq. $\eqref{tHooft formula}$ to compute the determinant.

An important aspect of the decomposition in Eq. $\eqref{decomposition rep R}$ is the relationship between the Dynkin index of the original $SU(N)$ representation and the sum of all the Dynkin indices of the corresponding $SU(2)$ representations. This is expressed as follows
\begin{align}
\textnormal{Tr}\left[T^{a}(\textbf{R})T^{b}(\textbf{R})\right]
=T(\textbf{R})\delta^{ab}= \sum_{i,j\rightarrow \textbf{R}}\textnormal{Tr}\left[\tau^{a}(t_i)\tau^b(t_j) \right]
=\sum_{i\rightarrow \textbf{R}} T(t_i)\delta^{ab}\, ,
\end{align}
where $T(\textbf{R})$ is the Dynkin index of the $SU(N)$ representation $\mathbf{R}$ and we used the linearity of the trace and the definition of the Dynkin index of an irrep $t_i$ of $SU(2)$.

\subsection{Complex scalar field charged under $SU(N)$}

When considering a single complex scalar field charged under an arbitrary representation $\textbf{R}$ of $SU(N)$, we can take advantage of the fact that $\mathcal{M}_{\phi}$ is block diagonal with respect to the decomposition into $SU(2)$ irreps. Consequently, we can apply Eq. $\eqref{tHooft formula}$ to each independent block of isospin $t_i$. This results in a factor of Eq. $\eqref{tHooft formula}$ for each block, leading to a contribution after Gaussian integration given by
\begin{align}
\left(\textbf{det} \mathcal{M}_{\phi}\right)^{-1}=&\exp\left[-\frac{1}{3}\sum_{i\rightarrow \textbf{R}} T(t_i)\ln(\mu\rho)-\sum_{i\rightarrow \textbf{R}}\alpha(t_i) \right]=\exp\left[-\frac{1}{3}T(\textbf{R})\ln(\mu\rho)-\sum_{i\rightarrow \textbf{R}}\alpha(t_i) \right]\, .
\end{align}
Note that we have obtained the complex scalar contribution to the $\beta$-function coefficient of the gauge coupling.

\subsection{Weyl spinor charged under $SU(N)$}

The background field expansion around the instanton solution gives us the following determiant over fermion non-zero modes\footnote{Recall that in the background of an instanton $\mathcal{M}_{\psi}^{(+)}=i\sigma_{\mu}D_{\mu}$ possesses $2T(R)$ zero modes, whereas $\mathcal{M}_{\psi}^{(-)}=i\bar{\sigma}_{\mu}D_{\mu}$ has none. Despite this difference in zero modes, both $\mathcal{M}_{\psi}^{(+)}\mathcal{M}_{\psi}^{(-)}$ and $\mathcal{M}_{\psi}^{(-)}\mathcal{M}_{\psi}^{(+)}$ share the same spectrum of non-zero modes \cite{Vandoren:2008xg}. For a detailed discussion of the fermion sector, see Appendix \ref{fermion zero modes and sources}.}
\begin{align}
\textbf{det'} \mathcal{M}_{\psi}\equiv & \left(\frac{\text{det'}\left(\mathcal{M}_{\psi}^{(-)}\mathcal{M}_{\psi}^{(+)}\right)}{\text{det}\left(\mathcal{M}_{\psi}^{(-)}\mathcal{M}_{\psi}^{(+)}+\mu^2\right)}\frac{\text{det'} \left(\mathcal{M}_{\psi}^{(+)}\mathcal{M}_{\psi}^{(-)}\right)}{\text{det'} \left(\mathcal{M}_{\psi}^{(+)}\mathcal{M}_{\psi}^{(-)}+\mu^2\right)}\right)^{1/4}\frac{\text{det} \left(\mathcal{M}_{\psi}^{0}+\mu\right)}{\det \mathcal{M}_{\psi}^{0}}\nonumber\\
=& \mu^{-T(\textbf{R})}\left(\frac{\text{det'}\left(\mathcal{M}_{\psi}^{(\pm)}\mathcal{M}_{\psi}^{(\mp)}\right)}{\text{det'} \left(\mathcal{M}_{\psi}^{(\pm)}\mathcal{M}_{\psi}^{(\mp)}+\mu^2\right)}\right)^{1/2}\frac{\det \left(\mathcal{M}_{\psi}^{0}+\mu\right)}{\det \mathcal{M}_{\psi}^{0}}\nonumber\\
=& \mu^{-T(\textbf{R})}\frac{\text{det'}\mathcal{M}_{\psi}}{\text{det'}(\mathcal{M}_{\psi}+\mu)}\frac{\text{det}(\mathcal{M}_{\psi}^0+\mu)}{\text{det}\mathcal{M}_{\psi}^{0}}\, .
\end{align}
We extracted $2T(\mathbf{R})$ factors of $\mu$ from the determinant in the denorminator, corresponding to the zero modes of $\mathcal{M}_{\psi}^{(-)}\mathcal{M}_{\psi}^{(+)}$, to obtain a dimensionless ratio of the primed determinants.
Since spinor space is two dimensional we have the relation $\left(\det\mathcal{M}_{\phi}\right)^2=\text{det'} \left(\mathcal{M}_{\psi}^{(+)} \mathcal{M}_{\psi}^{(-)}\right)$.
Considering a Weyl fermion charged under some representation $\mathbf{R}$ of $SU(N)$, we can again decompose its operator into blocks corresponding to $SU(2)$ irreps. We have again a factor of Eq. $\eqref{tHooft formula}$ for each of the block associated to the isospin $t_i$ involved in the decomposition. Thus, the fermionic contribution will have the form
\begin{align}
\textbf{det'} \mathcal{M}_{\psi}=&\mu^{- T(\textbf{R})}\exp\left[\frac{1}{3}\sum_{i\rightarrow \textbf{R}} T(t_i)\ln(\mu\rho)+\sum_{i\rightarrow \textbf{R}}\alpha(t_i) \right]=\rho^{ T(\textbf{R})}\exp\left[-\frac{2}{3}T(\textbf{R})\ln(\mu\rho)+\sum_{i\rightarrow \textbf{R}}\alpha(t_i) \right]\, .
\end{align}
Note that we recover the fermion contribution to the $\beta$-function coefficient of the gauge coupling.

\subsection{Pure Yang-Mills}

Now, we focus on the pure Yang-Mills sector of the generating functional. We need to evaluate the product of two determinants
\begin{equation}
\left(\textbf{det'} \mathcal{M}_A\right)^{-1/2}\left(\textbf{det} \mathcal{M}_{\rm ghost}\right)\, .
\end{equation}
Given that $\mathcal{M}_A$ possesses $4T(\mathbf{Adj}) = 4N$ zero modes, while $\mathcal{M}_{\rm ghost}$ has none, we extract the corresponding factors of $\mu^2$ from the determinant to obtain a dimensionless ratio of primed operators
\begin{equation}
\textbf{det'}\mathcal{M}_A=\mu^{-8N}\left[\exp\left(\frac{1}{3}\sum_{i\rightarrow \textbf{Adj}} T(t_i)\ln(\mu\rho)+\sum_{i\rightarrow \textbf{Adj}}\alpha(t_i) \right)\right]^4\, ,
\label{determinant gauge boson}
\end{equation}
where the power of $4$ comes from the relation between $\mathcal{M}_A$ and $\mathcal{M}_{\psi}^{(-)}\mathcal{M}_{\psi}^{(+)}$. From Eq. $\eqref{operators}$ we have
\begin{equation}
    \left(\mathcal{M}_{A}\right)_{\mu\nu} = \text{Tr}\left[\sigma_{\mu}\left(\mathcal{M}_{\psi}^{(-)}\mathcal{M}_{\psi}^{(+)}\right)\bar{\sigma}_{\nu}\right]\, ,
\end{equation}
which allows to establish that $\text{det'} \mathcal{M}_A = \left(\text{det'} \mathcal{M}_{\psi}^{(-)}\mathcal{M}_{\psi}^{(+)}\right)^2$, and therefore $\text{det'}\mathcal{M}_A=\left(\text{det}\mathcal{M}_{\phi}\right)^4$. The sum in the exponential of Eq. $\eqref{determinant gauge boson}$ is over the isospin representations involved in the decomposition of the generators of the adjoint representation of $SU(N)$. This is given by\footnote{This can be seen from the fact that the generators of the fundamental representation of $SU(N)$ decompose as
\begin{equation}
T^{a}(\mathbf{N})=\tau^{a}(1/2)\oplus(N-2)\tau^{a}(0)\, ,
\end{equation}
and since $\mathbf{N}\otimes \mathbf{\overline{N}}=\textbf{Adj}\oplus \mathbf{1}$ we obtain the desired decomposition.}
\begin{equation}
T^{a}(\mathbf{Adj})=\tau^{a}(1)\oplus 2(N-2)\tau^{a}(1/2)\oplus (N-2)^2\tau^{a}(0)\, .
\end{equation}
Thus, the gauge boson contribution is given by
\begin{equation}
\textbf{det'} \mathcal{M}_A=\mu^{-8N}\left[\exp\left(\frac{N}{3}\ln(\mu\rho)+\alpha(1)+2(N-2)\alpha(1/2) \right)\right]^4\, ,
\end{equation}
since $\alpha(0)=0$ and the Dynkin index of the adjoint representation is $T(\mathbf{Adj})=N$.
The contribution from the Faddeev-Popov ghosts is
\begin{equation}
\textbf{det} \mathcal{M}_{\rm ghost}=\exp\left[\frac{N}{3}\ln(\mu\rho)+4\alpha(1)+2(N-2)\alpha(1/2) \right]\, .
\end{equation}
Therefore, the pure Yang-Mills contribution to the one-loop determinant is given by
\begin{align}
\left(\textbf{det'} \mathcal{M}_A\right)^{-1/2}\left(\textbf{det} \mathcal{M}_{\rm ghost}\right)&= \rho^{-4N}\exp\left[\frac{11}{3}N\ln(\mu\rho)-\alpha(1)-2(N-2)\alpha(1/2) \right]\, ,
\end{align}
where the first term is the gauge fields and ghosts contributions to the $\beta$-function coefficient of the gauge coupling.
\newline

This completes the computation of the one-loop determinants for an $SU(N)$ gauge theory with any matter content.

\section{Fermion zero modes and sources}\label{fermion zero modes and sources}

We need to compute the generating functional in the background of an instanton to take into account interactions in a consistent way. To achieve this, we introduce sources for the field content, and in particular for fermions. In the instanton background, $\xi$ has no zero modes, while $\xi^{\dagger}$ possesses $2T(\textbf{R})$ zero modes, where $\textbf{R}$ denotes the representation of $\xi^{\dagger}$ under the gauge group. In this context, it is convenient to express the generating functional as follows
\begin{equation}
Z_0[\eta,\eta^{\dagger}]=\int\mathcal{D}\xi\mathcal{D}\xi^{\dagger}\exp\left[-\int d^4x \left(\frac{i}{2}\xi\sigma_{\mu}D_{\mu}\xi^{\dagger}+\frac{i}{2}\xi^{\dagger}\bar{\sigma}_{\mu}D_{\mu}\xi+\eta\xi+\xi^{\dagger}\eta^{\dagger} \right) \right]\, .
\end{equation}
We decompose $\xi$ in terms of the eigenfuctions $\upphi_i$ of $\mathcal{M}_{\psi}^{(+)}\mathcal{M}_{\psi}^{(-)}$ with Grassmann coefficients $b_i$, and $\xi^{\dagger}$ in terms of the eigenfunctions $\uppsi_i^{\dagger}$ of $\mathcal{M}_{\psi}^{(-)}\mathcal{M}_{\psi}^{(+)}$, with corresponding Grassmann coefficients $\bar{b}_i$. Among the latter, certain terms correspond to zero modes, for which we denote the Grassmann coefficients as $\bar{a}_i$
\begin{align}
\xi(x)=\sum_{\epsilon_i\ne 0}b_i \upphi_i(x)\, , \qquad\xi^{\dagger}(x)=\sum_{i=1}^{2T(\textbf{R})}\bar{a}_i\uppsi^{\dagger}_{0,i}(x) +\sum_{\epsilon_i\ne 0}\bar{b}_i \uppsi^{\dagger}_i(x)\, ,
\end{align}
Both $\mathcal{M}_{\psi}^{(+)}\mathcal{M}_{\psi}^{(-)}$ and $\mathcal{M}_{\psi}^{(-)}\mathcal{M}_{\psi}^{(+)}$ have the same spectrum of non-zero eigenvalues $-\epsilon_n^2$, and the associated eigenfunctions are related by
\begin{equation}
\upphi_n=\frac{1}{\epsilon_n}\mathcal{M}_{\psi}^{(+)}\uppsi^{\dagger}_n\, ,\qquad \uppsi_n^{\dagger}=-\frac{1}{\epsilon_n}\mathcal{M}_{\psi}^{(-)}\upphi_n\, .
\end{equation}
From this expression we see that $\upphi_n$ and $\uppsi_n^{\dagger}$ are orthogonal for different eigenvalues, and they have the same norm, given by
\begin{equation}
    \upupsilon_n\equiv \int d^4x\, \upphi_n(x)\upphi_n(x)\, ,\qquad \bar{\upupsilon}_n\equiv\int d^4x \,\uppsi_n^{\dagger}(x)\uppsi_n^{\dagger}(x)\, , \qquad \upupsilon_n=\bar{\upupsilon}_n\, ,
    \label{original definition of the norms}
\end{equation}
for non-zero eigenvalues.
Thus, plugging everything into the generating functional gives
\begin{align}
Z_0[\eta,\eta^{\dagger}]=&\left(\prod_{i=1}^{2T(\textbf{R})}\int\frac{d\bar{a}_i}{\sqrt{\bar{\upupsilon}_{0i}}}\right)\left(\prod_n\int \frac{db_n}{\sqrt{\upupsilon_n}}\frac{d\bar{b}_n}{\sqrt{\bar{\upupsilon}}_n}\right)\exp\left[-\int d^4x\left(\frac{1}{2}\sum_{n,m}b_n \bar{b}_m \upphi_n\mathcal{M}_{\psi}^{(+)}\uppsi_m^{\dagger}\right.\right.\nonumber\\
+&\frac{1}{2}\sum_{n,m}\bar{b}_n b_m \uppsi_n^{\dagger}\mathcal{M}_{\psi}^{(-)}\upphi_m +\left.\left.\sum_{n}\left[b_n\upphi_n\eta+\eta^{\dagger}\bar{b}_n\uppsi_n^{\dagger}\right]+\sum_{i=1}^{2T(\textbf{R})}\eta^{\dagger} \bar{a}_i \uppsi^{\dagger}_{0i} \right) \right]\, .
\label{generating functional 1}
\end{align}
The terms involving the non-zero modes become
\begin{equation}
\Bigg(\text{det'}\left(\mathcal{M}_{\psi}^{(-)}\mathcal{M}_{\psi}^{(+)}\right)\text{det'}\left(\mathcal{M}_{\psi}^{(+)}\mathcal{M}_{\psi}^{(-
)}\right)\Bigg)^{1/4}\exp\left[-\int_{x,y}\eta^{\dagger}(x)\cdot S'(x,y)\cdot \eta(y) \right]\, ,
\label{non-zero mode part of fermion}
\end{equation}
where $S'(x,y)$ is the Green's function of the operators $\mathcal{M}_{\psi}^{(\pm)}\mathcal{M}_{\psi}^{(\mp)}$, excluding the zero modes. For additional details, refer to Appendix B of \citep{Espinosa:1989qn}. The contribution from the zero modes is given by
\begin{equation}
\left(\prod_{i=1}^{2T(\textbf{R})}\int\frac{d\bar{a}_i}{\sqrt{\bar{\upupsilon}_{0i}}}\right)\exp\left[-\int_x \eta^{\dagger}(x) \uppsi^{\dagger}_{0}(x)\right]\, ,
\end{equation}
where $\uppsi_0^{\dagger}$ contains all the $2T(\textbf{R})$ fermion zero modes of $\xi^{\dagger}$.
By combining all the pieces, the free generating functional in the instanton background can be expressed as
\begin{equation}
Z_0[\eta,\eta^{\dagger}]=\left(\text{det'}\mathcal{M}_{\psi}\right)\exp\left[-\int_{x,y} \eta^{\dagger}(x)\cdot S'(x,y)\cdot \eta(y) \right]\left(\prod_{i=1}^{2T(\textbf{R})}\frac{d\bar{a}_i}{\sqrt{\bar{\upupsilon}_{0i}}}\right)\exp\left[-\int_x \eta^{\dagger}(x)\cdot \uppsi^{\dagger}_0(x) \right]\, ,
\label{Fermion generating functional instanton background}
\end{equation}
where to simplify notations we denote by $\text{det'}\mathcal{M}_{\psi}$ the combination of determinants in Eq. $\eqref{non-zero mode part of fermion}$. In what concerns us, the exponential term involving $S'$ will not contribute because we always set $\eta=\eta^{\dagger}=0$ at the end of the computations. As a result, in the background of an instanton, once a functional derivative with respect to $\eta^{\dagger}$ acts on this exponential, it will be eliminated by setting $\eta^{\dagger}=0$. Thus, we will not include this exponential factor, retaining only the $\eta^{\dagger}$ dependence of $Z_0$.

\bibliography{refs.bib}
%\bibliographystyle{utphys}

%%%%%%%%%%%%%%%%%%%%%%%%%%%%%%%%%%%%%%%%%%%%%%%%%%%%%%%

\end{document}